\documentclass[sn-mathphys-num]{sn-jnl}
\usepackage{paper-2024-simulation-bdi-jakta-alchemist}
\usepackage{natbib}

\begin{document}

\title{
    Testing \acs{BDI}-based \aclp{MAS} using \acl{DES}
}

\author[1]{\fnm{Martina} \sur{Baiardi}}\email{m.bariardi@unibo.it}
\equalcont{}

\author[1]{\fnm{Samuele} \sur{Burattini}}\email{samuele.burattini@unibo.it}
\equalcont{These authors contributed equally to this work.}

\author*[1]{\fnm{Giovanni} \sur{Ciatto}}\email{giovanni.ciatto@unibo.it}

\author[1]{\fnm{Danilo} \sur{Pianini}}\email{danilo.pianini@unibo.it}

\affil*[1]{
    \orgdiv{\disi},
    \orgname{\unibo},
    \orgaddress{
        \street{Via dell'Università, 50},
        \city{Cesena (FC)},
        \postcode{47522},
        \state{Emilia--Romagna},
        \country{Italy}
    }
}

\abstract{
Multi-agent systems are designed to deal with open, distributed systems
with unpredictable dynamics,
which makes them inherently hard to test.
The value of using simulation for this purpose
is recognized in the literature,
\firstReview{although} achieving sufficient fidelity
\firstReview{(i.e., the degree of similarity between the simulation and the real-world system)}
remains a challenging task.
This is exacerbated when dealing with \firstReview{cognitive agent models},
such as the \firstReview{\acf{BDI}} model,
\firstReview{where the agent codebase is not suitable to run unchanged in simulation environments,
thus increasing the reality gap between the deployed and simulated systems.}
We argue that \acs{BDI} developers should be able to
test in simulation \emph{the same} specification that will be later deployed,
with no surrogate representations.
Thus,
in this paper,
we discuss how the control flow of \acs{BDI} agents can be mapped onto a \acf{DES},
showing that such integration is possible at different degrees of granularity.
We substantiate our claims by producing an open-source prototype integration between two pre-existing tools
(\jakta{} and \alchemist{}),
showing that it is possible to produce a simulation-based testing environment for distributed \acs{BDI} agents,
\firstReview{
and that different granularities in mapping \acs{BDI} agents over \acsp{DES}
}
may lead to different
\firstReview{
degrees of fidelity.
}
}


\maketitle
\acresetall

\section{Introduction}\label{sec:introduction}

Modern software systems are increasingly complex,
and novel programming paradigms are emerging
to tackle \firstReview{application domains} such as
\acp{STS}\firstReview{~\cite{sociotech}}, \acp{CPS}\firstReview{~\cite{cps}}, or \acp{PS}\firstReview{~\cite{pervasive}},
where people interact with many computational entities situated in a physical environment
that feature automation, autonomy, and intelligence.
\firstReview{In these contexts},
\acp{MAS} have been widely adopted to tame complexity~\cite{Xie2017},
as they model software as a collection of interacting autonomous entities.

Depending on whether \acp{MAS} were exploited to \emph{study} systems' complexity,
or to \emph{tame} it,
\firstReview{many solutions have been proposed to either
use \ac{MAS} abstractions to \emph{simulate} system behaviour~\cite{BandiniMV09}
or for designing and \emph{deploying} real-world systems~\cite{CalegariCMO21}.}

The difference is subtle but crucial.
When simulated,
\firstReview{
    agents operate in a virtual sandbox environment
    designed to imitate the dynamics of the target real-world deployment
    in which
}%
perceptions (e.g., data from sensors, the flow of time)
and actions (e.g., agents' movements in the environment)
are controlled by the simulator,
\firstReview{and can be reproduced deterministically}.
When deployed, agents are attached to specific hardware and software runtimes,
communicate through networks, and are tied to the flow of real time.
\firstReview{
Agents could either run in virtual environments as concurrent
(and often distributed) processes
interacting with shared resources
(e.g., files, databases, network sockets)
hence perceiving the state of such resources in real time and acting on them.
In some cases, agents may even be embodied in physical entities
(e.g., robots, drones, embedded devices)
hence perceiving and acting through physical sensors and actuators.
}
Any of these elements may introduce non-determinism and unpredictability in the system,
to which the \ac{MAS} must be ready to react.

\firstReview{
Accordingly, in this paper, when referring to \emph{real-world deployment},
we intend software agents running on their final ``production'' platform
rather than within a simulation,
regardless of whether they are situated in a physical environment or not.
With \emph{real world} we therefore mean the target execution context for which the agents
are designed and implemented.
}

Aside from being useful in in-silico studies,
\emph{simulation} may also aid the \emph{development} and \emph{validation}~\cite{uhrmacher_simulation_2002}
of \acp{MAS} intended for real-world deployment.
In fact, simulation enables developers to test the behaviour of agents
and their dynamics in complex environments ahead of deployment,
while postponing \firstReview{and (ideally) reducing} costs \firstReview{and efforts, while mitigating} risks, associated with real-world execution.
%
%
The price for
\firstReview{having the possibility to test a \ac{MAS} specification in a simulated environment}
is paid in additional development effort.
\firstReview{F}irst and foremost, \firstReview{developers must} model the \emph{target} deployment context \firstReview{as} a simulation environment.
\firstReview{Moreover, developers must develop the behavior of the agents within the simulation framework, which may differ from the one used for real-world deployment, making sure}
to maintain alignment between the two versions of the \ac{MAS} codebase.

\firstReview{While the first challenge is unavoidable for any kind of simulation,
we argue that the second one could be mitigated in case that} the same \ac{MAS} specification can execute \emph{with no changes}
on both real hardware and a simulator of choice.

The situation is particularly challenging when dealing with \acp{MAS} featuring cognitive models of agency,
such as the \ac{BDI} model~\cite{RaoG95}.
\ac{BDI} agents have been proposed for complex scenarios, including
healthcare~\cite{CroattiMRGAA19},
multi-\acs{UAV} coordination~\cite{SilvestreLDBHB23},
social simulations~\cite{AdamGaudou2016},
and
cyber-security~\cite{NunesSF17}.
%
%
%
The abstraction gap between the \ac{BDI} programming model and most simulators~\cite{singh_integrating_2016},
along with the minimal support of \ac{BDI} technologies
for producing \emph{reproducible} simulations of articulated scenarios~\cite{kehoe2016robust}
lead developers to resort to one of the following approaches:
\begin{enumerate}
    \item extension of the \ac{BDI} platform with a dedicated simulation engine,
    which requires additional development effort;
    \item construction and maintenance of two parallel codebases
    (one for simulation and one for the actual system),
    leading to consistency issues;
    or
    \item integration of the \ac{BDI} platform with a general-purpose simulation engine,
    typically by synchronising their execution through some form of middleware~\cite{singh_integrating_2016}.
\end{enumerate}

All such approaches come with their own drawbacks,
and,
above all,
they are symptomatic of \firstReview{the lack of a general solution}.
In fact,
while the effort of modelling the environment is unavoidable for any simulation
-- and challenging per se~\cite{simchallenges} --,
porting agents' behaviour back and forth between simulation and real systems
should be as simple as possible,
ideally requiring no changes to the code.

\paragraph{Problem Statement}
In our view,
the state of the practice of \ac{BDI} systems engineering
lacks a general-purpose, practical solution for testing \ac{BDI} systems \firstReview{using} simulation,
before deployment.
Put differently,
switching between simulation and real-world deployment is currently cumbersome and impractical.
Ideally,
starting from the same codebase expressing a \ac{BDI} system specification,
developers should be able to test it in a simulated environment,
or run it on real hardware,
with no changes to the codebase,
while expecting the system to behave consistently in both scenarios.

The problem may seem merely technical,
but indeed reaching this goal requires redesigning how \ac{BDI} specifications are written and interpreted.
They should abstract away the implementation details of their execution,
while multiple execution engines should support both simulations and real-world deployments.
In the particular case of simulations,
the environment as well should be programmable,
other than \firstReview{only} agents.


\firstReview{
In practice,
we focus on a particular class of \ac{BDI} programming frameworks:
namely,
the ones implementing the \agentspeak{} language~\cite{RaoG95}%
---which gained widespread adoption among \ac{BDI} programmers
according to some recent surveys~\cite{computers10020016,CalegariCMO21}.
In the remainder of this work,
we will write ``\ac{BDI} framework'' to refer to \agentspeak{}-based \ac{BDI} frameworks,
unless explicitly stated otherwise.
}
\paragraph{Contribution}

In this paper,
\firstReview{
we address the long-standing problem of
\textbf{
enabling \ac{BDI}\firstReview{-based} \acp{MAS} to be executed in a simulation environment
with no (or minimal) changes to the original specification
}~\cite{uhrmacher_simulation_2002}.
We do so by tackling the following sub-objectives:
\begin{enumerate}[label=\textbf{O\arabic*}]
    \item\label{rq:methods}
    \Copy{RQ1text}{
        Investigate the practical trade-offs of the methods that have been proposed
        to develop and test \ac{BDI} agents in simulation.
    }
    \item\label{rq:abstract-spec}
    \Copy{RQ2text}{
        Understand which abstractions a \ac{BDI} framework should provide
        to integrate seamlessly with a simulation engine,
        such that the same specification can be executed in both simulation and real-world deployment.
    }
    \item\label{rq:des}
    \Copy{RQ3text}{
        Analyse how different mappings between \ac{BDI} agents' execution model and \ac{DES}
        can be designed, and how they impact the simulation fidelity.
    }
    \item\label{rq:sim-integration}
    \Copy{RQ4text}{
        Integrate a \ac{BDI} framework and a \ac{DES} simulation engine
        in such a way that the same specification can be executed in simulation and
        in real-world deployment, proving feasibility.
    }
\end{enumerate}
}

\paragraph{\firstReview{Strategy}}
Core to our contribution is the identification of \ac{DES} as a key enabler for this goal,
as well as an underexplored approach for \ac{BDI} systems simulation.
Along this line,
the first contribution of this paper is the
identification of critical challenges that emerge when
attempting to map a \ac{BDI} \ac{MAS} on a \ac{DES} simulator\firstReview{.}
%
\firstReview{
We analyze first the conceptual implications of such mapping,
focusing on the granularity of how \ac{BDI} agents' control loops can be mapped onto \ac{DES} events,
highlighting the trade-offs of different design choices.
Then, we discuss the technical implications of such mapping,
which impact the design of the agent programming framework,
to make it easier to integrate seamlessly with a simulator.
}

To substantiate our claims and to demonstrate the feasibility of our approach,
we also implement a prototype solution based on the integration of two existing tools,
namely:
\jakta{}~\cite{BaiardiBCP23} -- a \ac{BDI} programming framework --
and
\alchemist{}~\cite{PianiniJOS2013} general-purpose \ac{DES} simulator.
The key novelty here is that our prototype focuses on making \ac{BDI} simulations reproducible and \firstReview{transferable},
while also supporting different granularity levels and, therefore, different trade-offs in terms of \firstReview{complexity} and fidelity.
Aside from providing a practical tool for \ac{MAS} developers,
our goal is to document the experience gained during this integration.

Finally,
we exercise the prototype in a (simulated) multi-drone coordination scenario,
to exemplify how it can be exploited as means for the \emph{early} testing of \ac{BDI} systems.
\firstReview{More specifically},
we show how simulating the same \ac{BDI} specification with different complexity/fidelity trade-offs
may help in spotting issues that would have been otherwise overlooked.

\paragraph{Structure of the paper}

In \Cref{sec:background},
we recall most relevant concepts behind \ac{BDI} agents and simulation,
\firstReview{proposing \ac{DES} as a}
natural and robust way to simulate \ac{BDI} systems.
Then, in \Cref{sec:related-works},
we discuss existing approaches to simulate \ac{BDI}-programmed software in complex environments,
commenting on their limitations.
In \Cref{sec:bdi-over-des},
we show how to bridge the gap between \ac{BDI} and \ac{DES},
commenting on the technical implications simulating \acp{MAS},
and the granularity at which the \ac{BDI} reasoning cycle can be mapped into simulation events to increase simulation fidelity.
We then present in \Cref{sec:prototype} an open-source prototype that allows a \ac{BDI} agent specification
written in \jakta{}
to be executed on the \alchemist{} simulator.
By leveraging the modularity and extensibility features of the selected technologies
we are able to achieve a fully integrated solution
that does \emph{not} rely on ad-hoc synchronisation
mechanisms,
while still benefitting from the robustness of a state-of-the-art simulation engine.
Finally,
in \Cref{sec:evaluation},
we exercise the prototype in a multi-drone coordination scenario,
discussing how choosing different mappings between \ac{BDI} and \ac{DES} events can lead to misleading results,
before concluding and discussing future works in \Cref{sec:conclusion}.

\section{Background}\label{sec:background}

Here,
we recall the main notions behind \ac{BDI} agents and computer simulation,
under a unifying perspective rooted in the notion of \emph{event}.
Accordingly,
\firstReview{in \Cref{subsec:BDI-agents}} we \firstReview{first} summarise the state of the art of \ac{BDI} agents architectures,
\firstReview{then we analyse} which events are involved in their lifecycle.
We also report \firstReview{in \Cref{subsec:discrete-simulation}}
major definitions such as discrete-event, -time, and multi-agent-based simulation,
to ground our selection for the mapping of \ac{BDI} in simulation.

\subsection{\firstReview{The \acl{BDI} Model}}
\label{subsec:BDI-agents}

The \ac{BDI} model,
rooted in human psychology,
is based on the explicit representation of the cognitive process of
agents'
decision-making.
%
Originally a philosophical concept~\cite{BratmanEtAl1987},
it is now used to program~\cite{BordiniHW2007}
\firstReview{and simulate~\cite{HubnerB09} agents,
and model social behaviour~\cite{AdamGaudou2016}.}
\ac{BDI} systems are therefore particular cases of \acp{MAS},
where agents are endowed with \emph{cognitive} abstractions.

More precisely,
the model in \firstReview{\citet{RaoG95}} prescribes that every agent: 
\begin{inlinelist}
    \item memorises (possibly partial, or wrong) \emph{beliefs} about itself and its surrounding environment,
    that it can update by perceiving the environment, by reasoning, or by agent-to-agent communication;
    \item is driven by internal \emph{desires} (also known as ``goals'') it is willing to accomplish,
    the desires may change over time
    or lead to new desires when pursued;
    \item may commit to several concurrent \emph{intentions} as an attempt to pursue its desires;
    \item applies \textit{plans}, i.e., recipes of \textit{actions} that (the agent expects)
    will lead to the satisfaction of its desires.
\end{inlinelist}

\paragraph{\firstReview{\ac{BDI} \firstReview{Architectures}}}
\label{ssec:computational-bdi}

Many architectures have been proposed in the literature to implement \ac{BDI} agents in software,
including \agentspeak{}~\cite{RaoG95},
\ac{PRS}~\cite{IngrandGR1992},
and \ac{dMARS}~\cite{DInvernoLGKW04}.
In this work,
we focus on the former,
as it is one of the most widely adopted in the literature~\cite{CalegariCMO21,computers10020016}.

\agentspeak{}~\cite{RaoG95} agents are grounded on the notion of \emph{events},
to which agents react by selecting plans driving the execution of the agent.
Differently from purely reactive systems,
in which events are external stimuli directly mapped to actions,
\ac{BDI} agents are capable of reasoning about such events based on the context of their runtime state
and select plans accordingly.
\firstReview{
Plan execution can produce additional internal events
(e.g., sub-goals)
that will drive subsequent iterations of the agent's reasoning cycle.
}
Hence, for the remainder of this paper, when we refer to \ac{BDI} agents as \emph{event-driven},
we aim to highlight that,
as we recall in the remainder of this section,
the entire lifecycle of a \ac{BDI} agent is driven by events,
either received as environmental perceptions,
or generated by the internal processes of the agent.
Such event-driven nature of \ac{BDI} agents is a key feature,
especially when it comes to simulate them,
as \emph{in-silico} simulations essentially deal with the generation and processing of events
in a controlled and predictable way.
For these reasons,
in the remainder of this work,
we abuse the notation by referring to \ac{BDI} agents and \agentspeak{} agents interchangeably.

\paragraph{\firstReview{Events}}
\label{ssec:bdi-events}

\firstReview{All of the main abstractions of the \ac{BDI} model can be linked to events that drive the agent's behaviour.}
Beliefs \firstReview{-- encoding information currently stored in the agent memory and considered to be true -- are} updated in response to \emph{events} from the environment (perception), other agents (messages),
or by the agent itself during the execution of plans. 
When the agent adopts a new goal,
a new intention is created to choose a plan to pursue such new goal.
Plan executions may cause new events
(e.g., sub-goals, belief updates, actions on the environment)
to be generated, and so on.

\firstReview{We can distinguish two main categories of events: \emph{internal} and \emph{external}.
Such a distinction is not present in the \agentspeak{} model,
nor necessarily implemented by \ac{BDI} frameworks,
but it is useful in the remainder of this work to discuss the mapping of \ac{BDI} agents onto \ac{DES}.}

\firstReview{
One the one side,
\emph{internal} events are generated by each agent during the execution of its own control loop,
and they are meant to make it progress.
These include:
\begin{itemize}
    \item belief additions, removals, updates;%
    \item the addition or failure of novel (sub-)goals.
\end{itemize}
}

\firstReview{
On the other side,
\emph{external} events
are related to the environment and the changes therein occurring,
and, in particular,
how these changes are perceived or provoked or reacted to by the agents.
External events include -- but are not limited to -- an agent:
\begin{itemize}
    \item entering or exiting the environment (i.e., terminating);
    \item\label{item:percepts} \firstReview{perceiving} or receiving messages from the environment;
    \item sending a message to some recipient;
    \item executing an action that mutates the environment.
\end{itemize}
}

\firstReview{
In the current practice,
the specific way external and internal events are handled
may vary across \ac{BDI} implementations
and be fairly different between in-silico simulations and real-world executions,
indicating a substantial abstraction gap.
}

\firstReview{
Additionally, it is important to notice that
further \emph{external} events may occur independently of agent actions
and provoke changes in the environment.
These may or may not be explicitly modelled,
but for the sake of \ac{BDI} agents programming,
it is sufficient to model the \emph{perception} of these changes.
For instance,
should a \ac{BDI} system be affected by the environmental temperature,
a real-world deployment would require temperature sensors
\firstReview{(or an external temperature monitoring service)}
to be attached to the agents,
to perceive a \emph{spontaneously}-changing temperature from the environment.
Should the same \ac{BDI} system be simulated,
the simulator would be in charge of generating events related to temperature variations,
so that the agents can observe them and react accordingly.
}

\begin{figure*}
    \centering
    \includegraphics[width=\textwidth]{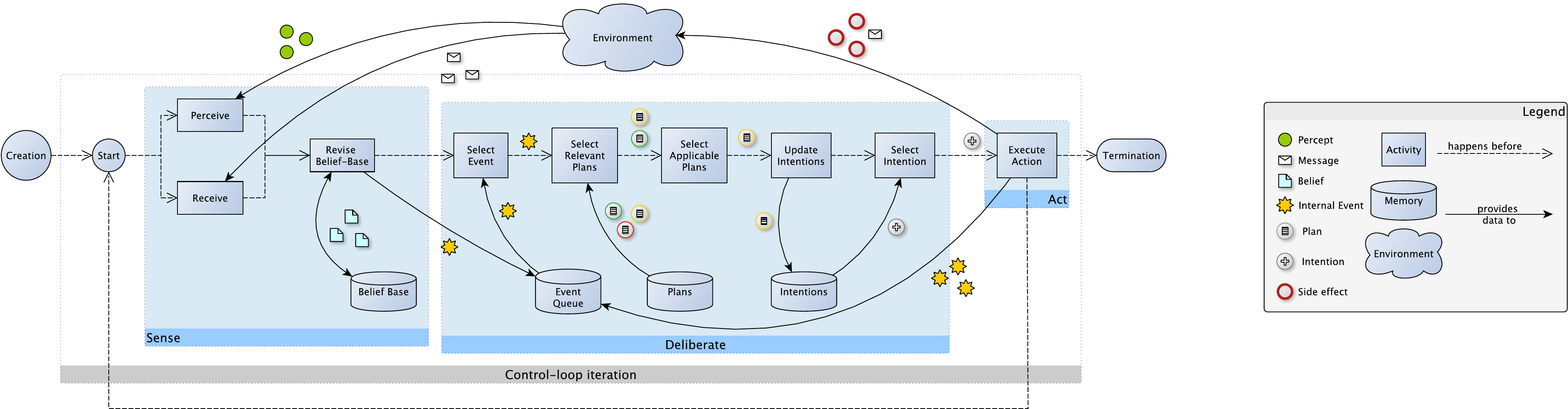}
    \caption{
        Graphical representation of the \agentspeak{} architecture for \ac{BDI} agents.
    }
    \label{fig:bdi-architecture}
\end{figure*}

\paragraph{\firstReview{Control Loop}}

Each \firstReview{\agentspeak} agent is animated by a control-loop,
i.e., a sequence of operations
(grouped into three major phases)
that are repeated indefinitely until the agent terminates.
Technically speaking,
the control-loop has been implemented in so many ways
-- each one coming with a different impact on the external concurrency of the agents~\cite{BaiardiBCPOR24b} --,
but the core abstraction is always the same.

\Cref{fig:bdi-architecture} shows a graphical representation of overall architecture of an \agentspeak{} agent,
with a focus on the major phases and operations composing the agent's control loop,
and the internal/external events therein involved.
Roughly speaking,
the control-loop assumes that each agent has a queue of internal events manipulated as follows.
Each iteration of the control-loop starts with the \emph{sense} phase,
where the agent collects
\begin{inlinelist}
    \item\label{ctloop:phase:sense1} percepts
    and
    \item\label{ctloop:phase:sense2} incoming messages,
    from the environment (i.e. external events),
    \item\label{ctloop:phase:sense3} updates the beliefs accordingly
    (enqueuing events from the updated beliefs).
\end{inlinelist}
Next, the agent executes the \emph{deliberation} phase, where it
\begin{inlinelist}[\alph]
    \item\label{ctloop:phase:delib1} picks one event from the queue;
    \item\label{ctloop:phase:delib2} selects a plan for handling the event,
    assigning it to some pre-existing or novel intention;
    and, consequently,
    \item\label{ctloop:phase:delib3} selects an intention to advance.
\end{inlinelist}
Finally, the agent executes the \emph{act} phase, where it
executes one action from the intention selected in step~\ref{ctloop:phase:delib3}.
The action execution may provoke further internal events,
which are then enqueued
(and therefore processed in the next iterations of the control-loop),
as well as external events,
which are delivered to the environment or other agents.

\paragraph{\firstReview{Execution}}

The execution of a \ac{BDI} agent boils essentially down to scheduling control-loop phases
and handling events.
Practically,
this may vary across implementations
and be fairly different between simulations and real-world executions.
\firstReview{
    Specifically, defining how the control loop is scheduled
    may lead to different emerging behaviours of the overall \ac{MAS} due to
    which interleaving of agent's control loops is allowed by the implementation.
}
On the one side,
in simulations,
the simulator is in charge of (reproducibly) virtualizing the flow of time.
In this case,
`executing the \ac{BDI} system' means to advance the simulation time
while deciding which agent(s) shall execute their control-loop phases next.
The goal here is emulate concurrency, while keeping the execution of the agents deterministic and reproducible.
On the other side,
in real-world deployments,
the agents are executed by the operating system,
the flow of time corresponds to the real-world time,
and the agents are scheduled by the operating system's scheduler,
which may introduce non-determinism in the execution of the agents.
In this case,
the goal is to balance agents execution onto the available computational resources,
and to exploit parallelism accordingly.


\subsection{Simulation \firstReview{Concepts}: \acs{DES}, \acs{DTS}, \acs{MABS}}
\label{subsec:discrete-simulation}

We \firstReview{call} simulation the process of replicating a real-world process
in a simplified and controlled fashion to gather insights about its behaviour.
Simulation can be performed by different means,
in this paper we refer to \emph{in-silico} simulation
(where the system model is described in software and executed on a computer)
of systems that feature temporal evolution.

\firstReview{
A general problem in simulation is how to ensure that the simulated system has
sufficient similarity with the real-world system it aims to replicate.
The degree of similarity is often referred to as the
\emph{fidelity} of the simulation~\cite{DBLP:conf/wsc/MoonH13}.
A related concept often used in software engineering
is the \emph{reality gap}~\cite{DBLP:conf/ecal/JacobiHH95}, which
refers to the discrepancies that may arise when transitioning
from a design to its real-world implementation.
}

\firstReview{
When using simulation as a validation tool,
the reality gap can be used to refer to the differences between the behaviour of the simulated system
and of the real-world system once deployed.
The concepts are obviously related:
increasing the fidelity of the simulation reduces the reality gap,
and a wide reality gap is often a symptom of low fidelity.
In this paper we will use both concepts
-- which are valid for all kinds of simulated systems --
to discuss the trade-offs of different design choices
when simulating \ac{BDI} \acp{MAS}.
}

\paragraph{\firstReview{Main Kinds of Simulation}}

Temporal evolution can be driven by events that happen and cause time to advance, in case of \acf{DES};
or by ``forced'' leaps forward (typically of a constant amount of time)
that cause the re-evaluation of the system state, in case of \acf{DTS}.
Different models of temporal evolution are suitable for different scenarios,
with \acp{DES} being generally capable of capturing more fine-grained dynamics
(for instance, stochastic chemistry~\cite{Cai2007}),
while \acp{DTS} being better at scaling and parallelisation when the phenomenon under study has a continuous nature in time
(for instance, n-body simulations~\cite{Richardson2000}).

Simulation frameworks that rely on the notion of agent to model their domain
are called \ac{MABS}~\cite{DrogoulVM02,BandiniMV09},
of which some prominent examples in the literature include
NetLogo~\cite{Sklar07},
 Mason\cite{LukeCPSB05},
 RePast~\cite{NorthCOTMBS13},
 \firstReview{GAMA~\cite{Taillandier2018}}.
 and Alchemist~\cite{PianiniJOS2013}.
\ac{MABS} are designed to support the modelling of complex systems in terms of independent agents,
with the goal to replicate complex phenomena for which other models
(e.g., \acp{ODE}) are unsuitable~\cite{DBLP:books/tf/09/DrogoulMF09,Edmonds00}.

As such,
\ac{MABS} are \emph{not} primarily designed to support \emph{testing of agent-based software}.
Indeed, they expose their own notion of agent,
typically simpler than agent abstractions devoted to programming distributed systems,
such as \ac{BDI} agents.
Using a \ac{MABS} framework to simulate a \ac{BDI} \ac{MAS} requires a significant effort
to either write a simplified version of the system to be tested
(at the price of losing fidelity)
or build an adaptation layer mapping the original \ac{BDI} program onto the \ac{MABS}' abstractions.

\paragraph{\firstReview{Simulation for \ac{BDI} \acp{MAS} Testing}}
In this work,
we are interested in leveraging simulation as part of the development toolkit of \ac{BDI} \acp{MAS},
specifically to test the system's behaviour in a controlled environment ahead of time,
to discover bugs,
stress-test the system in corner cases,
measure performance,
investigate transients,
and so on.
To do so,
it is paramount that the \ac{BDI} specification (i.e., the program code) exercised in testing via simulation
is \emph{the same} that will be later deployed in the real-world system,
or the price paid in fidelity will hardly be acceptable
(and, as a consequence, testing via simulation will lose its potential value).
Although not designed for this purpose,
we argue that \ac{MABS} are the tool that most closely resembles the needs of \ac{BDI} developers.
As previously introduced, however,
there is a substantial abstraction gap between the two models,
filling which is part of the contribution of this work.
From the point of view of the time model,
\acp{DES} appear the most natural choice to simulate \ac{BDI} agents,
as they are event-driven in nature as discussed in \Cref{ssec:computational-bdi}.
One of the contributions
of this work is showing how the control-flow events of \ac{BDI} agents can be mapped
onto \ac{DES} engine events,
and what are the implications of alternative design choices.

\firstReview{
In exploring the trade-offs, we also consider practical considerations
that affect how simulation supports software validation in early development stages.
In many scenarios,
simulation could be used to explore a cheap virtual prototype of the system,
even before a real counterpart exists~\cite{DBLP:journals/jos/FortinoN13,DBLP:journals/csse/FortinoGR05}.
As development progresses, the expected fidelity of the simulation should naturally increase, improving its usefulness as a validation tool.
Achieving higher fidelity, however, often requires more fine-grained models,
which come with greater computational cost.
This leads to an inevitable trade-off between accuracy, development time, and simulation duration.
Different phases of the engineering process may therefore call for different fidelity levels: fast, low-detail simulations to rapidly test ideas early on, and slower, high-detail simulations later when behaviour must be assessed more thoroughly.
}

\section{\firstReview{\ac{BDI} Agents Simulation Methods}}
\label{sec:related-works}

To achieve \ref{rq:methods},
in this section,
we investigate the state of the art of \ac{BDI} agents simulation.

The idea of testing \acp{MAS} dynamics via simulation
dates back to the early 2000s~\cite{uhrmacher_simulation_2002}.
However,
despite recognizing the relevance and complexity inherent in testing \acp{MAS}~\cite{miles_why_2010,nguyen_testing_2011},
and the existence of methodologies that include simulation for \ac{MAS} development~\cite{cossentino_passim_2008},
there are not many tools that effectively allow one
\emph{
    ``to execute agents as they are and to switch arbitrarily between execution in the real environment and the virtual test environment''
}\cite{uhrmacher_simulation_2002}.
The lack of dedicated simulation tools
is especially relevant for \ac{BDI} agents,
whose agents' lifecycle is richer compared to other models.

This seamless switching is the primary goal for our work\firstReview{~\cite{DBLP:conf/dsrt/Baiardi24}}.
In fact,
keeping the \emph{same} agent specification
ensures that simulation results are not artefacts induced
by the software translation into a simulation-compatible model.
Additionally,
if the specification is the same,
issues discovered at deployment time that were not evident during testing
will be at most imputable to the abstraction gap of the environment model.
\firstReview{
A recent work has tackled this problem for non-\ac{BDI} agents~\cite{Briola2025},
by developing a simulation environment for the JADE \ac{MAS} platform~\cite{DBLP:books/wi/BellifemineCG07}.
The work focuses on achieving portability of the same agent behaviour between simulation and real-world execution,
by implementing a simulation environment that mimics the JADE run-time environment.
We exclude this work from our comparison, as JADE agents are not \ac{BDI}-based.
}

%
Several attempts to use simulation as a testing tool for \ac{BDI} \acp{MAS} have been done in the past,
falling into the three main categories~\cite{singh_integrating_2016},
described below.

\newcommand{\relatedExtendFramework}{(1)}
\newcommand{\relatedExtendSimulator}{(2)}
\newcommand{\relatedMiddleware}{(3)}

\paragraph{\relatedExtendFramework{} Adding Simulation Features in \ac{BDI} Frameworks}

Agent frameworks can, in principle,
be modified so that the \ac{MAS} environment is replaced by a simulated one.
\firstReview{In practice, this operation may be difficult
because it depends on the \textit{concurrency model}~\cite{BaiardiBCPOR24} adopted by the framework
and how customizable the framework \firstReview{is}.}
Nevertheless,
some \ac{BDI} programming frameworks
include simulated execution.
For instance,
\jason{}~\cite{HubnerB09} allows a \ac{DTS} execution mode
with steps that execute one reasoning cycle for each agent,
and in~\firstReview{\citet{ricci_exploiting_2020}} is present a proposal to apply \ac{DES} to \jacamo{}.

\paragraph{\relatedExtendSimulator{} Implementing \ac{BDI} Features in Simulators}

As discussed in \Cref{subsec:discrete-simulation},
the agent model in \ac{MABS} simulators are simpler than \ac{BDI} agents,
thus,
their execution requires a reimplementation of the agent interpreter on top of the simulation engine.
\ac{BDI} extensions exist for several popular \ac{MABS}.
Some basic \ac{BDI} and FIPA~\cite{o1998fipa} communication libraries
have been implemented in NetLogo~\cite{Sklar07}
for educational purposes~\cite{sakellariou_enhancing_2008},
and
a \ac{BDI} layer has been implemented on top of the GAMA~\cite{DrogoulACGGMTVVZ13} platform
to support social simulations with cognitive models~\cite{TaillandierBCAG16}.
Other approaches rely on the \ac{DEVS} formalism to represent \ac{BDI} reasoning,
such as JAMES~\cite{uhrmacher1998agents}.

\paragraph{\relatedMiddleware{} Integration through Synchronisation}

Finally,
an approach is the integration of an existing \ac{BDI} framework with an off-the-shelf simulator.
This is typically achieved through some form of synchronisation,
either through inter-process communication
or shared memory.
A generic framework for such integration is proposed in~\firstReview{\citet{singh_integrating_2016}}
suggesting the adoption of a standard interface to map \ac{BDI} agents with \ac{MABS} agents;
and showing validation through different combinations of \ac{BDI} and \ac{MABS} platforms.
In~\firstReview{\citet{davoust_architecture_2020}},
instead,
a shared memory approach is used to make \jason{} agents interact with the simulated environment asynchronously,
reducing the time the simulation engine spends \emph{waiting} for the agent reasoning.

\paragraph{Remarks}

Categories \relatedExtendFramework{} and \relatedMiddleware{} would preserve the same code for the agent behaviour.
Category \relatedExtendFramework{},
besides requiring a considerable effort to build a simulation engine from scratch,
would hardly match existing state-of-the-art simulators
in terms of performance,
environment modelling,
visualization,
and data export tools.
Using strategy \relatedMiddleware{}, as we discuss in more detail in \Cref{sec:bdi-over-des},
can have important consequences on the reality gap with respect to the real-world execution.
Depending on how synchronisation is achieved,
assumptions on agent executions may be introduced implicitly,
possibly leading to unreliable results.
Simulating the behaviour of an existing \ac{MAS} following strategy \relatedExtendSimulator{}
requires to rewrite the agents' behaviour using the simulator's \ac{API}.
This is critical,
as it is hard to guarantee that the software will be the same,
and costly to realize and maintain.

In this paper,
we take an approach that sits in between these categories.
We do start from an existing \ac{BDI} framework and an existing simulator,
but instead of relying on some form of synchronisation,
we make the simulator run \ac{BDI} agents directly.
This is similar to the approach used to
support social simulations by integrating MASON~\cite{LukeCPSB05} with \jason{}~\cite{BordiniHW2007}
by incorporating the \jason{}'s interpreter within MASON
to directly execute agents written in \agentspeak{}~\cite{caballero2011eaai}.
%
%
\firstReview{The main difference of our proposal is in the scope}:
we do not mean -- as \firstReview{\citet{caballero2011eaai}} -- to build better social simulations using \ac{BDI} agent modeling,
rather,
 we want to provide a toolkit to test the behaviour of deployable \ac{BDI} software systems.
This difference in scope makes the execution of a complete reasoning cycle at each simulation step as done in~\firstReview{\citet{caballero2011eaai}}
not viable under the fidelity point of view,
as we will discuss in \Cref{sec:bdi-over-des} and demonstrate in \Cref{sec:evaluation}.

\section{\firstReview{Integrating \ac{BDI} and \ac{DES}}}
\label{sec:bdi-over-des}


\firstReview{
In this section
(and, more specifically, in \Cref{sec:portable-agent-specifications}),
we discuss how to allow}
for testing the behaviour of \ac{BDI} agents systems
in a simulation environment,
without changes to the codebase of the \ac{MAS} specification (\ref{rq:abstract-spec})\firstReview{.}
%
In particular,
as mentioned in \Cref{subsec:discrete-simulation},
we are interested in studying the exploitation of \ac{DES} to serve this purpose (\ref{rq:des}).
At the intuition level,
this should be possible,
as the events generated by the \ac{BDI} agent\firstReview{'s control loop}
can be scheduled in a \ac{DES} engine and interleave with all other simulation events.

In principle,
given a \ac{MAS} made up of
\begin{inlinelist}
    \item agents\firstReview{'} behavioural specifications,
    \item an environment abstraction to handle perceptions, actions, and communication,
    and
    \item an execution platform to run the system,
\end{inlinelist}
one could retain the agents specification and swap the remainder with a simulator
to achieve the desired \firstReview{objective}.
However, this task is \emph{deceptively simple}:
in the following subsection we will analyse how each of these elements
needs to be carefully designed to make such swap possible.

\firstReview{
Before proceeding,
let us clarify that,
in the remainder of this paper,
we specifically focus on agents whose behaviour is \emph{manually} specified beforehand by human developers
-- such as \ac{BDI} agents --
rather than \emph{learned} autonomously%
---e.g., via reinforcement learning.
In principle, though, similar considerations apply for learning agents,
provided that learning occurs \emph{online}\footnotemark
-- meaning \emph{during} the agent's interaction with the environment --
as opposed to \emph{offline} training%
---which occurs \emph{before} agents' execution.
}

\footnotetext{
    \firstReview{
        If an agent's behavioural specification includes some form of \emph{online} learning capabilities,
        (e.g., by implementing Q-learning~\cite{Watkins1992}),
        then
        the agent would be able to learn from its experience in the simulated environment
        as it would do in the real-world deployment.
        However,
        we acknowledge that modern approaches to reinforcement-learning
        (e.g., deep reinforcement learning~\cite{FigueiredoPrudencio2024})
        commonly rely on an \emph{offline} training phase,
        which leverages the simulation environment to generate training data and update the agent behaviour
        (i.e., the learned \emph{policy})
        accordingly.
        This training phase is outside the scope of this work,
        but given the final specification of the agent behaviour
        (e.g., a trained neural network implementing a policy),
        our assumptions discussed at the beginning of \Cref{sec:bdi-over-des} still apply.
        In fact,
        the learned policy would simply work in the target environment
        (no matter whether it is simulated or real-world),
        provided that the agents' perceptions and actions are properly abstracted
        and wired to the neural network's input and output layers.
    }
}

\subsection{Portable Agent Specifications}
\label{sec:portable-agent-specifications}

\firstReview{
    We have established that the main goal of this work is to allow
    the same \ac{BDI} agent specification to be executed
    both in simulation and in real-world deployments.
    The technical implication of this goal is to ensure that the agent code
    is \emph{portable} i.e., it depends solely on platform-agnostic \acp{API},
    which can be implemented differently depending on the execution context
    (simulation or real-world).
}

Platform-specific \firstReview{implementations}
-- at least, one for real-world deployments and another for simulation --
must be provided for such a platform-agnostic \ac{API}.
\firstReview{
We refer to these implementations
as the \emph{interpreters} of the \ac{BDI} agent specification,
which are responsible for executing the agent code
by relying on the underlying platform functionalities.
}
The two interpreters may of course differ in their implementations,
yet both should guarantee the \agentspeak{} semantics is preserved.
In this regard,
a good practice is to let the two interpreters share as much code as possible:
ideally implementing the agent's control loop in a shared library,
which allows for plugging in different platform-specific functionalities.

\firstReview{
    Given the role of environment abstractions
    for \ac{MAS} programming~\cite{weyns2007aamas}
    that can be used to abstract agents from the execution context%
    -- ranging from distributed systems~\cite{IssicabaRPB17} to robotics~\cite{pantoja2016atal,moro2022paams} --%
    we discuss what are the fundamental requirements
    to design an environment \ac{API}
    that allows for portable \ac{BDI} agent specifications
    that can seamlessly run on a simulation platform.
    Namely, the environment \ac{API} must be designed to abstract the following functionalities.
}

\paragraph{Time}
Agent behaviour may depend on the passage of time.
Hence, agents may need to observe the current time,
usually through the system's clock in a real-world deployment.
In simulation, however, the flow of time is virtualized and controlled by the simulator,
and it usually does not correspond to real-world time.
Thus, access to \emph{time-dependent} functionalities
(e.g., timestamps, timeouts, delays)
must not be implemented by accessing system-level time primitives directly.
Accessing the system's clock, or using sleep functions
must be avoided in the agent code,
and instead the environment \ac{API} must provide the necessary functionalities
to access the current time
or to schedule timeouts and delays.
It will then be the responsibility of the integration with the simulation platform
to provide a proper implementation of such functionalities
upon the simulator's notion of time,
guaranteeing semantic correspondence with the real-world counterpart.

\paragraph{Randomness}
Sometimes,
the agents' behaviour may require a degree of \emph{non-determinism}%
---e.g., using random number generators.
In simulation,
to ensure reproducibility of simulation,
all non-determinism must be controlled by the simulator.
Thus,
similar to time-related primitives,
random number generation
must be provided by the environment \ac{API}.
Reproducibility is especially significant when simulations are a mean
to debug anomalies.
When debugging, any two executions with the same random seed
\emph{must} produce the same sequence of events,
otherwise, the original issue may not re-occur,
or, worse, may manifest differently,
making understanding and fixing the problem much harder---%
an issue known in the software engineering community as \emph{flaky behaviour}\cite{DBLP:journals/tosem/ParryKHM22}.

\paragraph{Perceptions and Actions}
Agents' interaction with the environment
happens through perceptions and actions.
On the one side, the environment \ac{API} must provide functionalities
to perceive the state of the environment entities and deliver perception \emph{events}
to the agents.
On the other side, the environment \ac{API} must encapsulate all the actions
that agents can take to influence the environment state.
This level of separation is the most common in \ac{MAS} programming frameworks,
but it is worth stressing its importance
when the goal is to run the same agent specification
both in simulation and in real-world deployments.
In (\Cref{ssec:design-simulated-environment})
we will discuss in depth the implications of this abstraction when
discussing how to bind the \ac{BDI} environment \ac{API}
to a \ac{DES} simulation platform.

\paragraph{Communication}
Agents in a \ac{MAS} often need to communicate with each other.
To support inter-agent
communication, the environment \ac{API} must provide functionalities to send and receive messages.
Real-world (distributed) deployments rely on network protocols
(e.g., HTTP, MQTT, XMPP, etc.),
and although in principle they can be used in simulation as well,
a typical choice driven by performance and improved experimental control
is to abstract them away and provide direct control over relevant control variables
(latency, message/package loss, retransmissions).
Typically,
in \ac{MAS}
the environment acts as a \emph{relay} for messages,
and messages might be subject to some sort of \emph{propagation} mechanism.
The way messages are propagated may depend on the scenario,
for instance,
taking into account the environment topology by
limiting the delivery of messages to \emph{reachable} destinations,
or
introducing some form of \emph{delay}
or \emph{loss}
in the delivery of messages
to better match with the expected real-world environment dynamics and
stress the properties of the system under different conditions.
Again, failing to abstract communication functionalities
would make impossible to simulate the system,
as the simulator would need to rely on the underlying network stack to deliver messages between agents,
which would make the simulation results unreliable and non-reproducible.

\subsection{Design of the Simulated Environment}
\label{ssec:design-simulated-environment}

\firstReview{
Having defined the requirements for portable \ac{BDI} agent specifications (\Cref{sec:portable-agent-specifications})
by encapsulating all platform-specific functionalities
into a generic environment \ac{API},
we now focus on the implications of implementing such \ac{API}
on top of a \ac{DES} simulation platform.
These considerations are generally valid for any simulation environment
that aims to support the typical application scenarios of (\ac{BDI}) \acp{MAS},
i.e., distributed systems where agents interact with each other
and with a dynamic environment, possibly featuring
entities that can move in a physical space.
Although not specific to \ac{BDI} agents,
we still discuss these aspects,
as they set requirements and constraints in the choice of the simulation platform
and in the design of the simulation environment
that will host the \ac{BDI} \ac{MAS} under test.
}

\firstReview{
The most important difference between real-world deployments and simulations
concerns the way \emph{external} events are delivered to the agents.
In this section we focus on the modeling of such events,
leaving the discussion on
the scheduling of \emph{internal} events to \Cref{ssec:granularity}.
}
The \ac{BDI} model already assumes agents to be reactive to the external events they perceive.
In real-world deployments,
developers can wire \ac{BDI} agents' perceptions and actions to physical sensors and actuators
\firstReview{or connect the agents with I/O peripherals and external services}.
External events just occur, and agents either perceive or cause them.
Instead,
in simulation,
external events are delivered to the agents by the simulator,
and therefore require additional modelling and programming efforts.


\paragraph{\firstReview{Topology and Situatedness}}

\firstReview{
\ac{BDI} \acp{MAS} can be deployed either in physical or virtual environments.
Depending on the nature of the target scenario,
}
to accurately replicate the real-world mechanisms,
the simulation environment must support the notion of \emph{situatedness} (cf. Wooldridge \cite{Wooldridge00}),
that is crucial for all kinds of agents
and determines the range of observability and influence of the agents.
In some cases,
for instance when the target scenario features robots or mobile devices,
agents need to be further \emph{embodied}~\cite{brooks1991ijcai},
making them directly part of the environment,
and observable to other agents.

While in real-world deployments the space and topology of a \ac{MAS} is a \emph{constraint},
in simulation they are aspects to be modelled and programmed;
and the programming efforts might vary significantly depending on the target application.
When agents are expected to be deployed on mobile devices,
the simulation can use manifolds
(e.g., 3D spaces or city maps)
where location and distance can be defined.
In other cases,
for instance in \emph{networked} systems,
manifolds are not relevant,
but the \emph{network topology} is.
These are \emph{logical} environments,
in which entities are located into nodes of a graph,
and distances are computed as functions of the paths connecting them.
In some applications,
these aspects mix:
for instance,
a wireless sensor network may feature geographically \emph{situated} devices
for which a logical topology can model the communication channels
(e.g., based on the relative distance of the devices).
In any case,
a notion of \emph{closeness} is necessary
to limit the perception range and influence of agents,
whose specific reification
should be left to the simulation implementation,
as it
is scenario-dependent.

\firstReview{
    These aspects are important to consider when implementing the environment \ac{API}
    on top of a \ac{DES} platform.
    For instance, the perception range influences which events
    should deliver perceptions to the agents,
    and the topology may affect how messages are propagated in the environment.
    These parameters should be carefully modelled in the simulation
    and can be used as variables to explore different simulated scenarios.
}

\paragraph{\firstReview{Environment Dynamics}}

In any non-trivial scenario,
there will be events generated externally to the \ac{BDI} agents,
as real-world environments are typically dynamic
(i.e., they change over time)
or feature some form of stochastic behaviour
(e.g., package loss in networked systems).
Movement,
communication,
and, more generally,
any phenomena that span over time affecting the agents' perceptions,
must be modelled into events with appropriate time distributions.
These events must be modelled in simulation
and properly interleave with the agents' reasoning cycle.
The interleaving between agents' internals and other events
essentially determines how agents are executed in the simulation,
as we discuss in the next section.
\firstReview{
The design of the environment dynamics
allows one to implement the \ac{API} used by the agents to
perform actions that have effects on the environment state,
but also to model external sources of change
completely independent of the agent behaviour,
that may challenge the \ac{MAS} execution,
thus stressing its properties
in different scenarios.
As for any simulation, the more accurate the model of the environment dynamics,
the higher the fidelity of the simulation will be.
}

\subsection{Mapping the \ac{BDI} Execution on a Simulator}
\label{ssec:granularity}

\begin{figure*}
    \centering
    \includegraphics[width=0.8\textwidth]{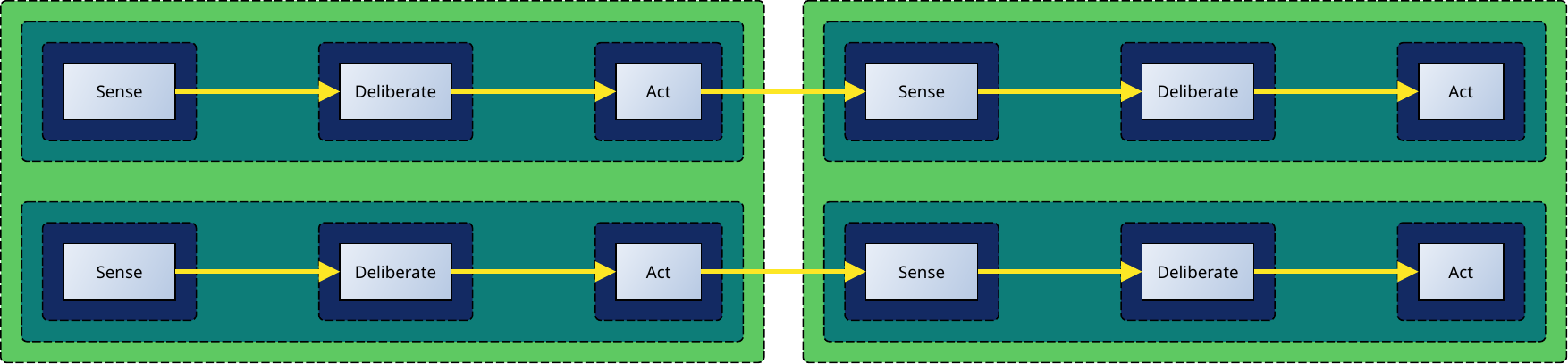}
    \caption{
        Granularities for atomic \ac{BDI} events.
        Yellow arrows are causal links.
        In \acs{AMA} (light green) an event is a run of all control loops of the whole \ac{MAS};
        it implicitly synchronises all agents.
        In \acs{ACLI} (emerald) an event is a single control loop iteration of a single agent,
        preventing phase interleaving.
        In \acs{ACLP} (dark purple), single control loop phases of every agent can interleave.
        \firstReview{\acs{ABE} is not shown as it is indistinguishable from \acs{ACLP} at the agent level.}
    }
    \label{fig:granularity}
\end{figure*}

Mapping \ac{BDI} events in a \ac{DES}
\firstReview{
to achieve \ref{rq:des}
}
requires an understanding
of which events are \emph{atomic} in a \ac{BDI} system execution.
Then,
any collection of these atomic events can be mapped to a single \ac{DES} event,
defining the mapping \emph{granularity},
and thus defining which \ac{BDI} groups of events can interleave with other simulation events.
The granularity influences the \emph{fidelity} of the simulation,
and is similar to choosing the external concurrency model~\cite{BaiardiBCPOR24} of a \ac{BDI} system upon deployment.
The granularity does not affect the individual agent program semantics
(the specification of behaviour is the same),
but may influence the collective dynamics of the \ac{MAS}
due to the interleaving of individual agent actions.
\firstReview{Coarse-grained mappings (i.e. multiple agents events mapped to one simulation event) introduce implicit synchronisation that suppresses possible interleavings, thus potentially hiding timing-dependent faults, race conditions, message races, and sensor (or actuators) delays that would happen in a real deployment.
Finer-grained (i.e. one agent event mapped to one simulation event)
mappings expose asynchrony and delays,
increasing fidelity and reducing the reality gap,
but at the cost of a more difficult integration mapping.
\Cref{fig:granularity} summarises the options that we discuss below,
showing the trade-offs for each choice.}

\subsubsection{\acf{AMA}}\label{subsubsec:ama}

Each \ac{DES} event corresponds to a \emph{full} control-loop iteration of \emph{all} agents in the system,
making it the most coarse-grained option.
This approach is adopted in~\firstReview{\citet{HubnerB09}} and~\firstReview{\citet{caballero2011eaai}},
which use \ac{DTS},
and it is similar to the one proposed in~\firstReview{\citet{singh_integrating_2016}},
which advances the simulation only after all agents
are \emph{idle}.
One obvious consequence of this choice is that
all agents in the \ac{MAS} run at the same \emph{frequency},
thus inducing implicit synchronisation,
and discarding all the effects caused by possible
\emph{interleaving} agents' actions.

\subsubsection{\acf{ACLI}}\label{subsubsec:acli}

Each \ac{DES} event corresponds to a \emph{full} control-loop iteration of a \emph{single} agent,
making this option slightly more fine-grained than \ac{AMA}.
In this case,
arbitrary interleaves among agents' \emph{actions} are possible,
but not among different agents' control-loop \emph{phases}.
This option does not allow modelling different durations for different phases of the control loop,
for instance,
it cannot be captured a situation in which
deliberation takes too long and makes the sensing phase outdated before an action is taken.

\subsubsection{\acf{ACLP}}\label{subsubsec:aclp}

Every single \emph{atomic phase}
(sense, deliberate and act)
is represented as a \ac{DES} event.
This option preserves the behaviour of concurrent or distributed \ac{BDI} agents,
allowing phases to interleave across different agents,
thus capturing complex inter-agent concurrency patterns.

\subsubsection{\ac{ABE}}\label{subsubsec:abe}

Each \ac{BDI} event is mapped to a single \ac{DES} event.
This is the finest-grained option,
first proposed in~\firstReview{\citet{ricci_exploiting_2020}}.
The approach has value for investigating the internals of the agent execution platform implementation,
e.g., to verify the correctness of the \ac{BDI} interpreter implementation
and its degree of parallelism,
but,
from the point of view of an observer of the \ac{MAS},
this model and \ac{ACLP} are indistinguishable,
as the phases of the control loop of each agent are atomic.
\firstReview{
    For this reason, in the remainder of the paper,
    we will consider \ac{ABE} to be subsumed by \ac{ACLP}.
}

\subsection{\firstReview{Tool Requirements for the Integration}}
\label{ssec:requirements}
\firstReview{Following the key features outlined in the sections above,
the tools to be adopted for the integration must provide:
\begin{itemize}
    \item neat \emph{modularisation} of the \ac{BDI} execution engine,
    to simplify its replacement with the simulation engine;
    \item abstract \emph{``external concurrency''} model~\cite{BaiardiBCPOR24},
    to allow for different granularities of the \ac{BDI} events
    -- and therefore different fidelity levels --
    in simulations;
    \item \emph{extensibility} of the \ac{DES} framework,
    which must accommodate the \ac{BDI} framework's abstractions with no major changes;
    \item \emph{simulation model} featuring locality, position, and communication channels,
    to support simulating agents situated in physical or logical spaces;
    \item native \emph{integration} (i.e., same runtime) between the \ac{BDI} framework and the simulator,
    to avoid the additional burden of inter-process communication mechanisms.
\end{itemize}}

\section{Prototype: \jakta{} over \alchemist{}}
\label{sec:prototype}


To demonstrate the feasibility of the ideas discussed in the previous sections,
and \firstReview{in order to achieve} \ref{rq:sim-integration},
here we present a proof of concept implementation of a \ac{BDI} interpreter supporting both real-world and \firstReview{simulation} execution.

In particular,
our discussion revolves around the \jakta{} \ac{BDI} framework~\cite{BaiardiBCP23},
which \emph{already} supports real-world execution of \ac{MAS} \firstReview{as concurrent applications},
and it allows for plugging in different \emph{concurrency models}~\cite{BaiardiBCPOR24b}.
\firstReview{Put simply,
\jakta{} is a \ac{BDI} interpreter which lets developers customise the way agents' control loop iterations -- as well as agents' perceptions and actions -- are executed.}
For the sake of conciseness,
in this section,
we only show how \jakta{} can be extended to support one more way to run agents,
namely:
in a simulated world,
as created by the \alchemist{} simulator~\cite{PianiniJOS2013}.
The interested reader can refer to~\firstReview{\citet{SNCS:JaktaJournal}} for a more detailed discussion on \jakta{},
and its support to \ac{BDI} agents in real-world deployments.

Further motivations exist for the technological choice of \jakta{} and \alchemist{},
as we discuss \firstReview{in \Cref{ssec:requirements} and} in \Cref{ssec:choices}.
As the reader will notice,
our prototype is tailored on these two technologies,
and the discussion in this section is specific to them.
However,
we believe that our experience has a general value:
not only it proves that \ac{BDI} systems' execution can be swapped between real-world deployments and \ac{DES} simulation
without changing the \ac{MAS} specification,
but it also allows us to study the impact of the many simulation granularity levels (see \Cref{ssec:granularity})
on the validation of a \ac{BDI} system.
This is indeed the purpose of \Cref{sec:evaluation}.

Technically speaking,
our prototype is implemented
as a stand-alone module in the \jakta's codebase,
named \texttt{alchemist-\allowbreak{}jakta-\allowbreak{}incarnation},
which imports simulator's \ac{API}
and allows developers to plug their \ac{BDI} specification
to be executed in a simulated environment.
The source code is publicly available\footnote{
    \url{https://github.com/jakta-bdi/jakta}
}
under a permissive licence
and archived on Zenodo~\cite{zenodojakta}
for future reference.


\subsection{\firstReview{Design and Technological Choices}}
\label{ssec:choices}
\firstReview{In line with the requirements in \Cref{ssec:requirements},
w}e selected the \alchemist{} simulator
and the \jakta{} \ac{BDI} framework
also because they both run on the \ac{JVM} and are therefore interoperable.
\alchemist{} is designed with extensible base abstractions,
thus allowing for reasonably straightforward integration with other stand-alone tools;
indeed,
it has been used in the past to simulate programmable tuple spaces~\cite{ZambonelliPMC2015},
biochemical systems~\cite{MontagnaSAC2012},
and aggregate programming languages~\cite{PianiniSAC2015,scafi}.
Complementarily,
\jakta{} has been designed to be easily adaptable to different execution engines.
So,
bridging the two technologies is a matter of implementing a new \emph{incarnation} in \alchemist{}
that can run \jakta{} agents.

\paragraph{\firstReview{\alchemist{}}}

\firstReview{\alchemist{}~\cite{PianiniJOS2013} is a general-purpose \ac{DES}
originally conceived to simulate bio-inspired systems,
from which it inherits the metaphors and terminology in use to this day.
In the remainder of this discussion,
terms in \texttt{typewriter font} represent names of existing software entities,
whose name has been used verbatim, as found in the \ac{API}.}

\firstReview{In \alchemist,
the simulated model entry point is the \texttt{Environment},
which represents a Riemannian manifold\footnote{A Riemannian manifold is
a generalised Euclidean space that can be curved
(useful to support geospatial data),
discontinuous
(useful to model inaccessible areas),
and for which the distance between two points $\overline{AB}$ may depend on the direction: $\overline{AB}\neq\overline{BA}$
(useful to implement asymmetric navigation, e.g., on street maps with one-way roads).}.
The \texttt{Environment} is populated by \texttt{Node}s, which
\begin{inlinelist}
    \item can be equipped with arbitrary \texttt{NodeProperty}s
    \item have a \texttt{Position} in space,
    \item can be programmed with \texttt{Reaction}s,
    \item can be connected to other \texttt{Node}s via a \texttt{LinkingRule} to form \texttt{Neighborhood}s,
    \item can contain \texttt{Molecule}s (data identificators),
    each with an associated \texttt{Concentration} (data value).
\end{inlinelist}}

\firstReview{
    Events in \alchemist{} are called \texttt{Reaction}s,
    as they were originally conceived to simulate chemical reactions.
    \texttt{Reaction}s in \alchemist{} extend the concept of reaction from chemistry.
    Instead of being defined by a set of reactants that
    transform into products at a rate defined by the \emph{law of mass action},
    \texttt{Reaction}s in \alchemist{} are more generally defined as
    collections of conditions, that, when satisfied,
    trigger a set of actions according to an arbitrary \emph{rate equation}
    which depends on the observable state of the \texttt{Environment}
    and a provided \texttt{TimeDistribution}.
}

\firstReview{
    The concept of \texttt{Concentration} in \alchemist{} can be reified
    in arbitrary data types.
    Different types of \texttt{Concentration} may lead to a different semantics
    for \texttt{Molecule}s, \texttt{Condition}s, \texttt{Action}s, \texttt{Reaction}s,
    and any other abstraction depending on them,
    thus allowing for different \emph{data models},
    and providing a flexible and straightforward way to
    bridge \alchemist{} with other tools.
    Every set of abstractions defining a coherent semantics
    for these entities is called an \texttt{Incarnation}
    in the simulator jargon.
    At the time of writing,
    \alchemist{} provides several incarnations
    for different application domains,
    including
    bio-chemical systems~\cite{MontagnaSAC2012},
    networked tuple spaces\cite{ZambonelliPMC2015},
    and aggregate programming via Protelis~\cite{PianiniSAC2015} and Scafi~\cite{scafi}.
}

\firstReview{
    An \alchemist{} simulation is configured using a YAML file,
    which specifies the \texttt{Environment} type,
    the \texttt{Node}s to populate it with,
    their initial \texttt{Molecule} concentrations,
    the \texttt{LinkingRule} that connects them,
    and any \texttt{Reaction}s to be executed.
    When the simulation starts,
    \alchemist{} loads the configuration file and instantiates
    the specified \texttt{Environment} and its contents.
}

\paragraph{\firstReview{\jakta{}}}
\label{par:jakta}

\firstReview{\jakta{}~\cite{SNCS:JaktaJournal} is a Kotlin-based modular \ac{BDI} framework.}
\firstReview{Each agent in a \jakta{} \ac{MAS} is equipped with an \texttt{AgentLifecycle} which implements the \ac{BDI} control loop.
\jakta{} cleanly separates the definition of the agents' behaviour from the execution platform
in charge of executing the agents' control loop phases,
thus allowing different concurrency models for agents to be plugged in.
\jakta's environments (not to be confused with \alchemist{}'s \texttt{Environment}s)
are lightweight abstractions that handle
perceptions, communications, and interaction with the external world.
\acp{MAS} in \jakta{} are programmed through a Kotlin \ac{DSL},
which provides an expressive, type-safe, and
\ac{IDE}-friendly way to define agents' beliefs, goals, and plans.
Indeed, by leveraging Kotlin as the host language,
\jakta{} programs are regular Kotlin applications,
which may lower the entry barrier for developers familiar with mainstream programming languages
(Kotlin is the reference language for Android development\footnote{
    ``Android’s Kotlin-first approach'', archived 2024-02-15 \url{https://archive.is/EtWqn}
})
and are thus natively supported by any \ac{IDE} supporting Kotlin,
significantly lowering the maintenance cost.
}

\paragraph{\firstReview{Multiple granularity support}}

As discussed in \Cref{ssec:granularity},
the mapping between the \ac{BDI} framework and the \ac{DES} simulator
can be realised with different granularity.
We implemented multiple granularities in our prototype
to show how each of them
impacts on the tested behaviour of a \ac{BDI} software.
We implemented all those introduced in \Cref{ssec:granularity} but \ac{ABE} because,
as previously discussed,
its behaviour is indistinguishable from \ac{ACLP}
\firstReview{
and supporting it
while retaining real-world deployment capability for \jakta{} specifications
would have been highly impractical\footnotemark.
}
\footnotetext{
    \firstReview{
        To support \ac{ABE},
        the \jakta{} interpreter should be modified
        to expose each atomic event as a schedulable unit of execution.
        While this is feasible in principle,
        the effort required is not justified by the benefits,
        as \ac{ABE} does not provide any additional value
        with respect to \ac{ACLP}.
    }
}

\paragraph*{}

Driven by these choices,
in the next subsection,
we proceed to identify the modelling abstractions of the two technologies
and we describe how they have been mapped
in order to merge the \jakta{} framework into the \alchemist{} simulator.

\subsection{\ac{BDI} \acp{MAS} in \alchemist{}}\label{subsec:mapping}

\begin{figure}
    \centering
    \includegraphics[width=\linewidth]{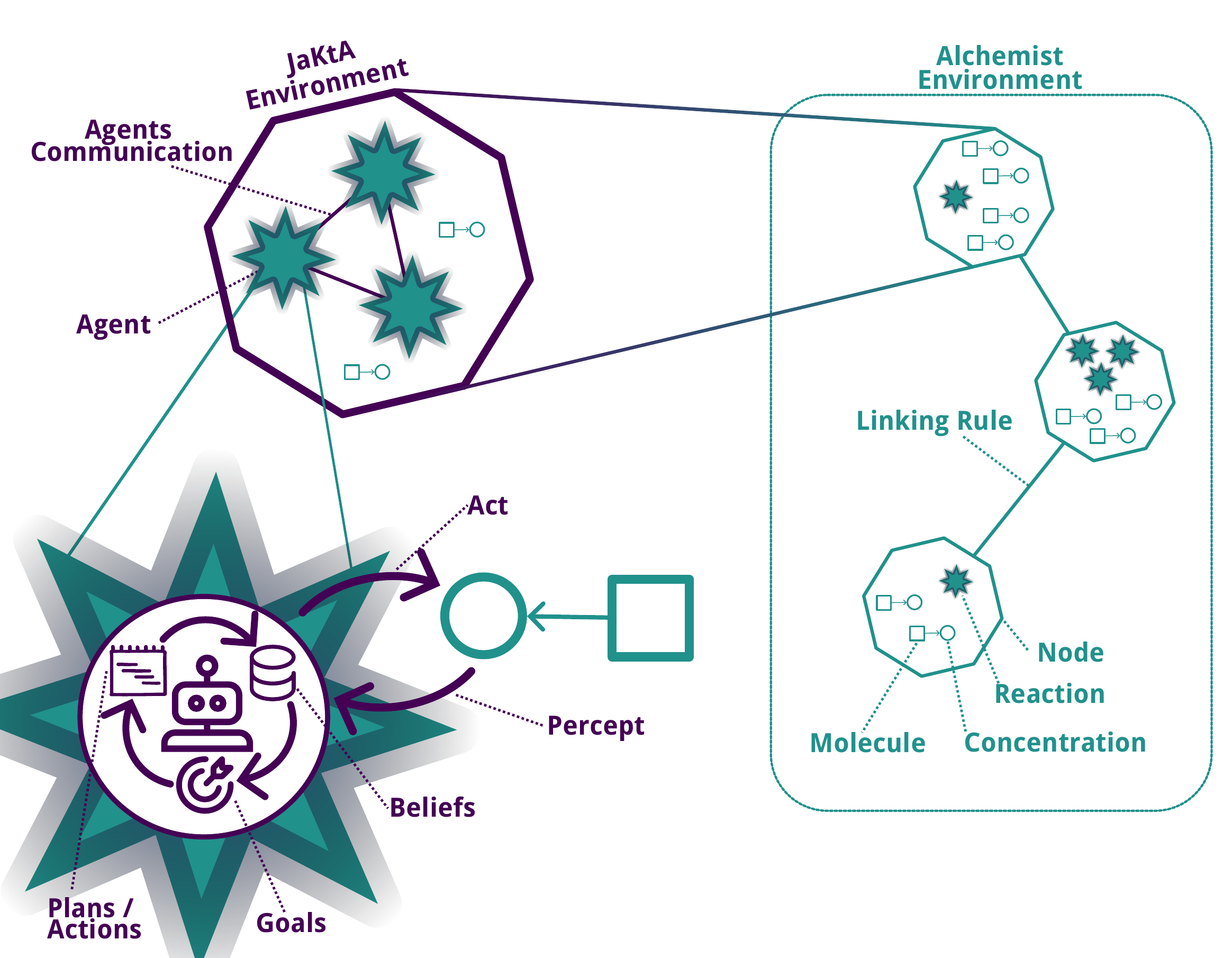}
    \caption{
        Mapping \jakta{} BDI abstractions onto \alchemist{} ones.
        Each \alchemist{} Node represents a physical \emph{device}
        running a centralised \jakta{} \ac{MAS}.
        Simulation of distributed \ac{MAS} is achieved having multiple Nodes in the same environment that communicate via \alchemist{}'s Linking Rules.
        Each agent's control-loop is implemented as an \alchemist{} \texttt{Reaction},
        with a custom \texttt{TimeDistribution} for each phase.
        Agents can act on and perceive \alchemist{}'s molecules in the environment.
    }
    \label{fig:mapping}
\end{figure}

\firstReview{
Integrating the two frameworks has been first an exercise of model-to-model mapping.
}
The mapping of the frameworks' abstractions on each other is summarised in \Cref{fig:mapping}.

Each agent's lifecycle phases \firstReview{(i.e., sense, deliberate, and act)} are mapped
onto separate \ac{DES} events (\alchemist{}'s \texttt{Reaction}s).
\firstReview{
As introduced in \Cref{ssec:choices},
each such event has a duration modelled through a \texttt{TimeDistribution}.
To guarantee that each agent's control loop phases are executed in order,
we implemented a custom \texttt{TimeDistribution}
(\texttt{JaktaTimeDistribution})
composed of three sub-\texttt{TimeDistribution}s,
one for each phase of the control loop.
\texttt{JaktaTimeDistribution} maintains an internal state
which tracks the current phase of the agent,
and executes the next phase according to its configured \texttt{TimeDistribution}.
This design allows us to flexibly support different granularities:
\begin{itemize}
    \item \ac{ACLP} is immediately and natively supported,
    as any three distributions can be picked for the three phases.
    \item To support \ac{ACLI},
    the \texttt{TimeDistribution} of the \emph{deliberate} and \emph{act} phases
    must be an as-soon-as-possible distribution
    (for instance, a negative exponential distribution with rate $\lambda=\infty$),
    thus allowing the next agent's control loop to be scheduled immediately after the previous one.
    \item To support \ac{AMA},
    in addition to the constraint used for \ac{ACLI},
    the \texttt{TimeDistribution} of the \emph{sense} phases across all agents
    must be scheduled to occur exactly once
    before any agent can execute a new control loop iteration:
    we do so by using, as \emph{sense} \texttt{TimeDistribution} for any agent,
    a \texttt{DiracComb} with a period $T$,
    starting at a random time $\tau_0 < T$.
    This strategy guarantees that every $T$ units of simulated time,
    all agents will have completed exactly one reasoning cycle,
    executed in some arbitrary (but deterministic and equal for each $T$ interval) order.
\end{itemize}
}

\firstReview{
While the \texttt{TimeDistribution} for internal events depends on the chosen granularity,
external events must be explicitly modelled in the simulation configuration.
For example,
the movement of a \texttt{Node} in the environment is an external \texttt{Event},
thus it must be specified with its own \texttt{TimeDistribution} in the simulation.
}

\texttt{Node}s in \alchemist{} naturally map onto situated devices,
as they possess a defined location in space.
Consequently,
we could not map \jakta{}'s \texttt{Agent}s directly onto \alchemist{}'s \texttt{Node}s,
as in a distributed \ac{BDI} \ac{MAS} multiple agents can be hosted on the same device.
Rather,
we mapped \alchemist{}'s \texttt{Node}s into instances of devices executing the \jakta{} platform,
thus capable of hosting multiple agents at once.

Mapping \texttt{Node}s to platform instances (or devices) enables direct inheritance of the
\firstReview{\alchemist{}'s} \texttt{Linking Rule} abstraction to model the communication channels,
and in principle allows for the implementation of mobile agents.
The Alchemist-compatible platform instance \firstReview{(named \texttt{JaktaEnvironmentForAlchemist})} is a \jakta{}-specific \texttt{NodeProperty},
adapting \jakta{} to run inside every \alchemist{} \texttt{Node},
driven by the simulator's control flow.

\firstReview{
    Agents can reason on the current device status,
    as information stored inside \texttt{Nodes}'
    is reified as perceived knowledge from \jakta{} agents.
    Information inside nodes is stored as key-value pairs
    (\texttt{Molecule}s) and associated \texttt{Concentration}s (c.f. \Cref{ssec:choices}),
    that agents can manipulate through actions.
    Changes are then treated by the simulator as ``regular''
    environment modifications, consequent to simulated events.
}

\firstReview{
    Communication among agents is modeled via an \alchemist{} \texttt{NodeProperty};
    each \texttt{Node} (device) is equipped with a message broker
    that collects messages for the agents on that device.
    When an agent sends a message to another agent on the same \texttt{Node},
    the message is delivered directly to the recipient's message queue.
    When the recipient is on a different \texttt{Node},
    the message is delivered if the recipient's \texttt{Node} is in the neighborhood
    of the sender's \texttt{Node},
    according to the \texttt{LinkingRule} defined in the simulation configuration.
}

\section{Evaluation}\label{sec:evaluation}

In this section,
we exercise the prototype introduced in \Cref{sec:prototype}
to demonstrate \emph{feasibility} of the proposed mapping of \ac{BDI} agents over \ac{DES}
and the \emph{transparency} of execution with respect to the agent specification.
Additionally,
we measure the impact granularity has on the \emph{reality gap} of the simulation \firstReview{($\ref{rq:des}$)},
showing that the same \ac{MAS} logic can produce different emerging behaviours
\firstReview{when executed with different granularities.}

\firstReview{
Our evaluation mimicks a situation where
multiple \acp{UAV}, each controlled by a \ac{BDI} agent,
must cooperate to maintain a circular formation.
%
%
To demonstrate the portability of the \ac{MAS} codebase
between simulation and real-world,
we show that the same code can be executed inside the simulator
or directly as a regular multithreaded application.
In the former case, the \ac{UAV} controls are simulated through the tools provided by the
\ac{DES} \ac{API},
while, in the latter case,
the \ac{UAV} controls are encapsulated in a different \ac{API}.
}

\footnotetext{
    \firstReview{
        One may argue that ``mocking'' is, in a broad sense, a form of simulation.
        However,
        as we discuss extensively in \Cref{subsec:discrete-simulation},
        we consider as ``simulation'' only the execution of a system in a
        \emph{virtualised} execution environment,
        where both time and space are virtual and detached from the real world,
        and where the whole dynamic of the system is controllable.
        This is not the case for mocked components.
    }
}

\firstReview{
It is worth highlighting that our experimental design
(specifically: the first part, where the \ac{MAS} is validated via \ac{DES})
is particularly useful to show the impact of simulation granularity onto validation.
In fact,
as we discuss in the next subsections,
choosing a coarser granularity for the simulation may hide potential issues in the agent programs,
which would instead emerge when a finer granularity is used.
}

\subsection{\firstReview{Use Case Description}}
\label{ssec:use-case}
\firstReview{
    We use a \acp{UAV} coordination scenario as our reference.
We use \jakta{} to instruct a flock of \acp{UAV},
each controlled by a \ac{BDI} agent,
to build and maintain a circular formation while following a moving leader \ac{UAV}.
The \emph{leader} \ac{UAV} moves in a circular path (radius $r_l=5m$),
    while follower \acp{UAV} must coordinate to build a circular formation around the leader
    so that every follower is at distance $r_f=2.5m$ from the leader,
    and all followers are equally distanced to each other.
    This formation must be maintained while the leader moves along its path.
}

\firstReview{We assume direct \ac{LoS} \ac{UAV}-to-\ac{UAV} communication within a short range of $r_c=5m$.
\acp{UAV} can navigate the space at a maximum speed of $1\nicefrac{m}{s}$.
If no signal is received from the leader, followers hover at their current location.
In this scenario, we assume \acp{UAV} are equipped with some localisation system that provides them with exact coordinates of their position in a given shared reference system.
Initially,
    follower \acp{UAV} are randomly displaced into a circular arena of radius $10m$,
    while the leader is located at the arena centre.
}

\firstReview{
    Each \ac{UAV} runs the \ac{BDI} control loop in rounds
    at a certain maximum execution frequency
    $f = \nicefrac{1}{T}$, cf. \Cref{subsec:mapping}.
Each control-loop phase would take some time to complete, depending on the complexity of the computations involved.
For instance, a frequency of $f=1Hz$ implies the execution
    of one agent control loop per second.
Tuning this frequency is a way to balance the responsiveness of the agent to changes in the
    environment with resource utilization (e.g., battery consumption),
and must be tuned to stay within the physical limits of the \ac{UAV}'s hardware.
}

\firstReview{
Additionally,
devices are not started simultaneously,
each of them experiences a random start delay $\tau_0 \in [0, \nicefrac{1}{f}]$
    (or, equivalently, $\tau_0 \in [0, T]$ see \Cref{subsec:mapping}).
}

\subsection{\firstReview{Modeling \ac{UAV} Behaviour in \jakta{}}}
\label{ssec:bdi-behaviour}

\firstReview{
During the experiment we deliberately used simplified \ac{BDI} programs that rely on tight synchronisation of agents' control loops.
We model two agent roles: a single \emph{leader} and multiple \emph{followers}.
}
\firstReview{The leader continuously traverses a circular trajectory at constant speed. }
\firstReview{Its behaviour is expressed as a single \ac{BDI} goal named \texttt{move}
(i.e., following the circular path)
and an associated plan that repeatedly:
\begin{enumerate}
    \item moves the leader to the next waypoint along the circle, and
    \item broadcasts a message to nearby agents (within communication radius $r_c$) inviting them to join the formation.
\end{enumerate}}
\firstReview{Followers do not possess a proactive goal;
they are reactive agents driven by messages from the leader.
Upon receiving a ``join formation'' message,
a follower executes a plan that computes and moves to its assigned position in the circular formation.
The target position is determined from the current number
of followers already in the formation.}
\firstReview{The \ac{BDI} program that expresses this behaviour is shown in \Cref{fig:code}.}

\subsection{\firstReview{Experimental Setup}}
\label{ssec:experimental-setup}
\firstReview{Given the shared behavior implementation described above,
in this section we describe how we experimented with the developed prototype
to demonstrate that:
\begin{itemize}
    \item different granularities in the simulation execution may lead to different emergent behaviours of the \ac{MAS},
    \item the same \ac{MAS} codebase can be executed both in simulation and on a real-world deployment without modification.
\end{itemize}}

\firstReview{
    \acp{UAV} in a real system would run independently.
    This makes the \ac{ACLP} granularity the most realistic one,
    as it allows for control-loop phase interleaving.
    Simulating at a coarser granularity
    may lead to incorrect forecasts about the correctness of the agent program.
}

\firstReview{
    To demonstrate the portability of the \ac{MAS} codebase to a real-world deployment,
we execute the \ac{UAV} control \ac{MAS} as a regular process
where all agents run concurrently as threads.
This is, of course,
    a simplified approximation of a deployment on real \acp{UAV},
    but it suffices to show that the same codebase can be executed
    without modification in both simulation and real-world settings.
}

\firstReview{
    We ensured code portability
    by enforcing that all interactions with the environment are done through the \jakta{} environment \ac{API},
    which we implemented both for the real-world deployment and for the \alchemist{} simulator.
}

\subsubsection{\firstReview{Simulated System Deployment on the \alchemist{} Simulator}}
\label{ssec:simulator-execution}

\firstReview{We use the simulator to compare the execution of the same \ac{MAS} logic
with different granularities, namely,
    we consider \ac{AMA}, \ac{ACLI}, and \ac{ACLP} (cf. \Cref{ssec:granularity}).
We expect that coarser-grained granularities show better system performance
than more finer-grained
(in our case, better formation maintenance).
The experiment is designed to show that such effect is 
a consequence of the over-simplification of the \ac{BDI} program
and of its execution using coarse-grained simulation.
%
}

\paragraph{\firstReview{Modeling \acp{UAV}}}

\firstReview{\acp{UAV} are modelled as \alchemist{} \texttt{Node}s.
Each \texttt{Node} is mapped to a \jakta{} platform instance
that hosts a single \jakta{} agent controlling the \ac{UAV} as described in \Cref{subsec:mapping}.}

\firstReview{Nodes can move in a 2D continuous space using the \alchemist{}'s built-in \texttt{MoveToTarget} action. Agents compute the desired destination based on the perceived positions of the other \acp{UAV} and the leader, and write it to a dedicated \texttt{Molecule} in the environment, which is then read by the \texttt{MoveToTarget} action during its execution which applies the movement considering a speed parameter.
This mechanism is used to make movement realistic and not instantaneous in the simulation, as \acp{UAV} can only move at a maximum speed of $1\nicefrac{m}{s}$.}

\begin{figure*}
    \centering
    \includegraphics[width=.32\linewidth]{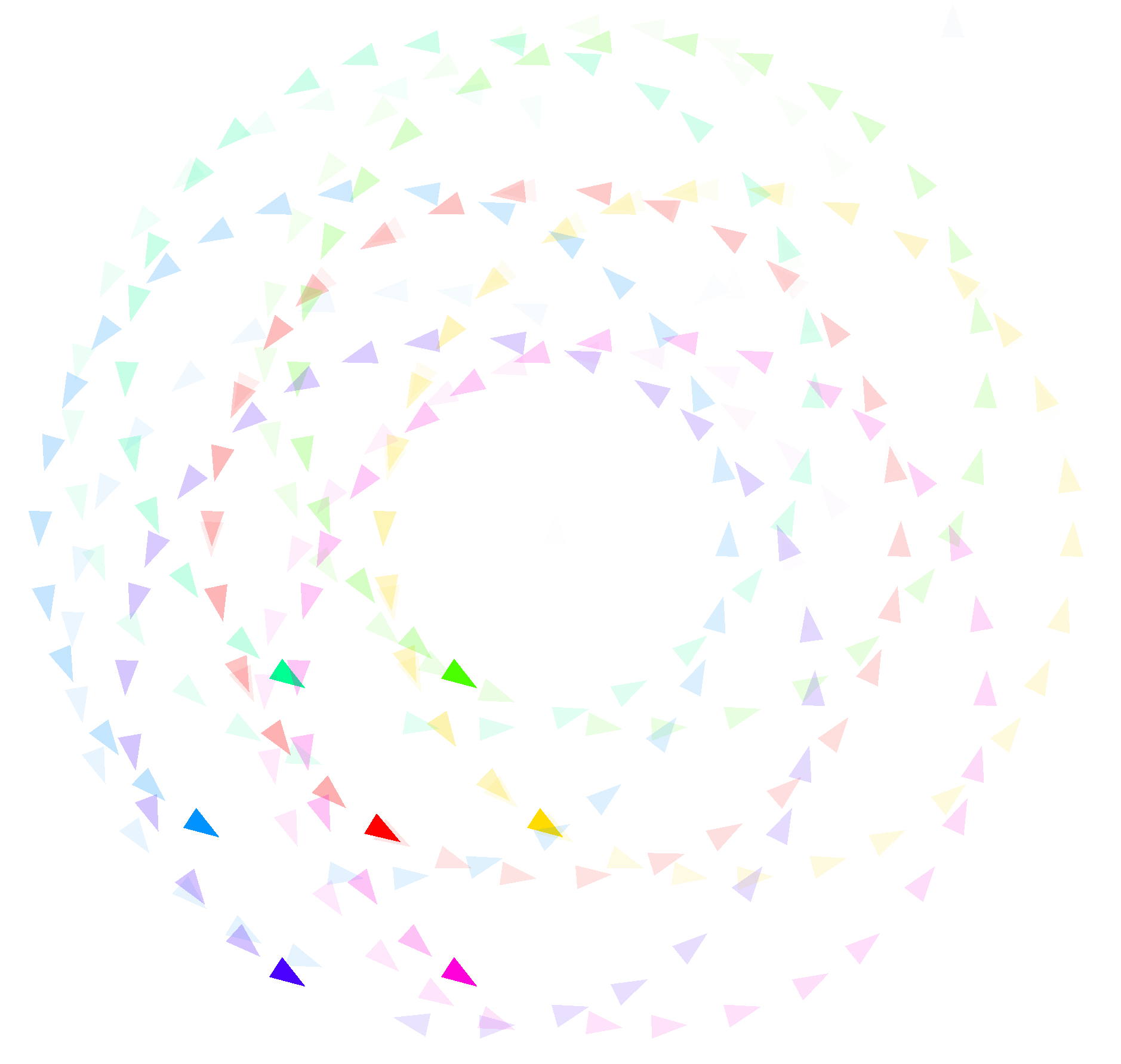}
    \includegraphics[width=.32\linewidth]{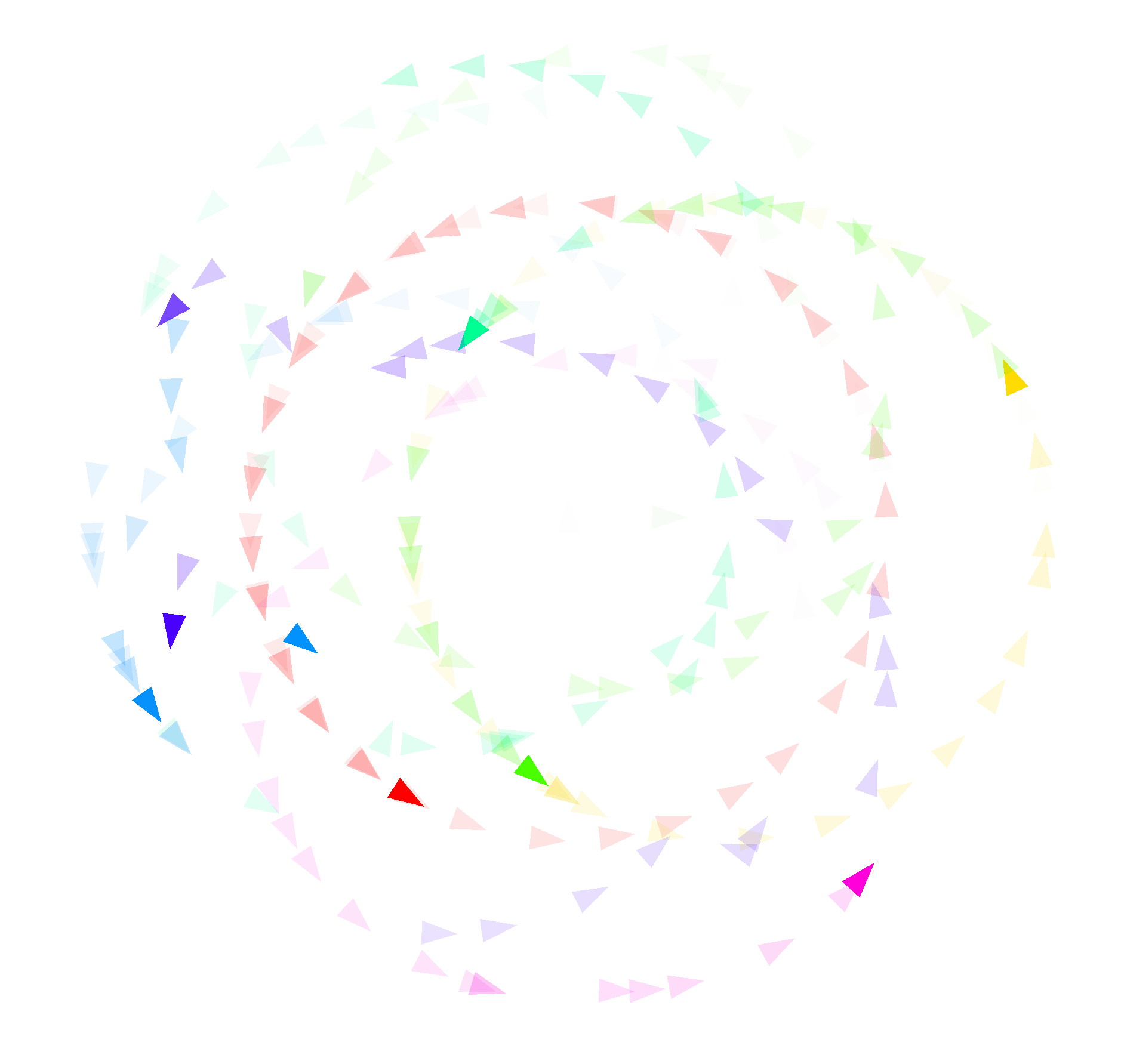}
    \includegraphics[width=.32\linewidth]{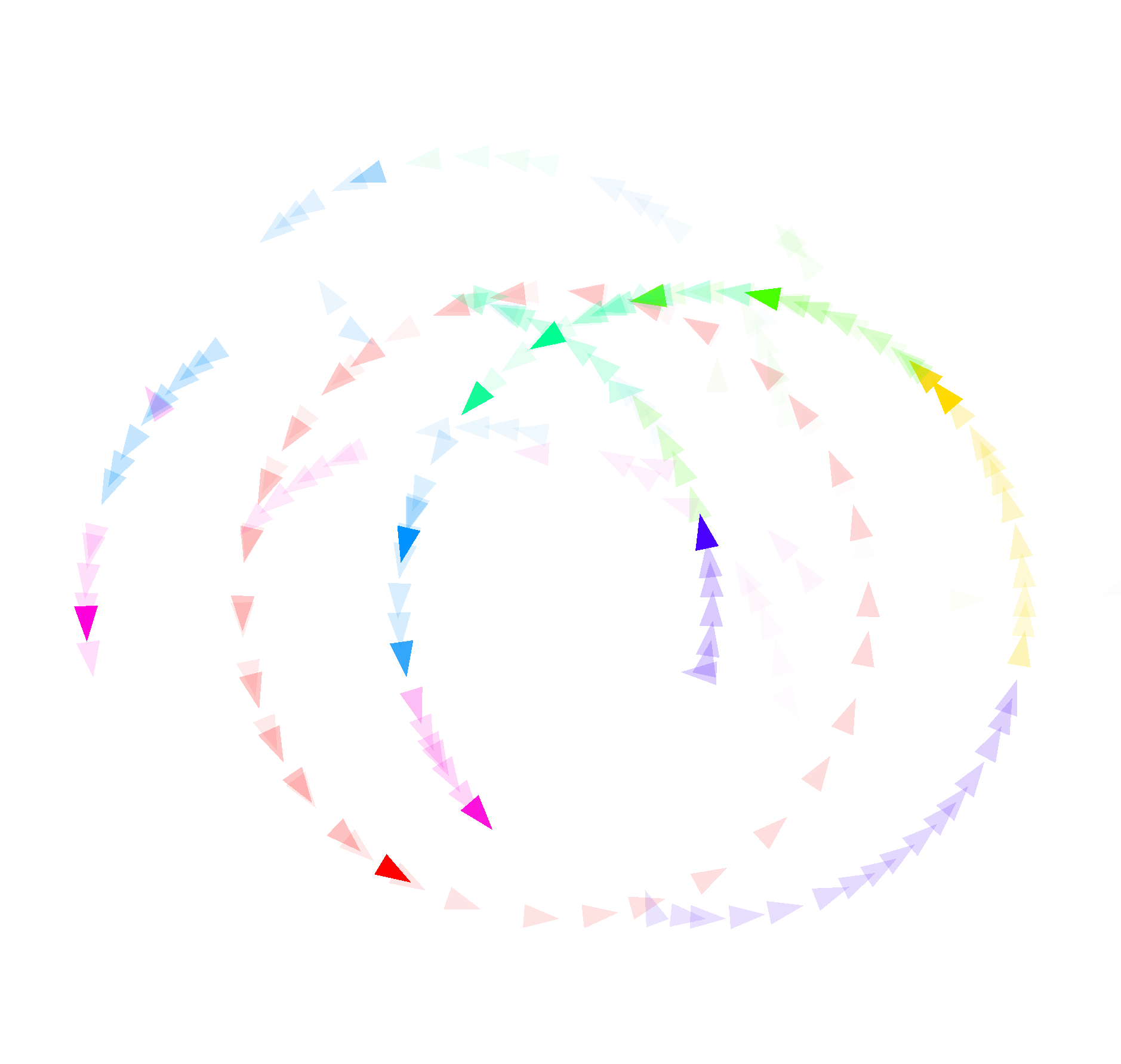}
    \caption{
        Simulation snapshots for the three cases: \ac{AMA} (left), \ac{ACLI} (centre), and \ac{ACLP} right
        for $\tau=0.6$.
        Every \ac{UAV} is depicted with a different colour, the leader is the red one.
        The follower \ac{UAV} count has been reduced to six for better visualisation.
        Color intensity captures the distance in time, more intense colors are closer to the current time.
        In \ac{AMA}, all \acp{UAV} follow a circular trajectory intersecting the leader's one.
        The formation is much less regular in \ac{ACLI}, and it is completely lost in \ac{ACLP},
        showing that failing to capture the model nuances correctly may produce systems that work
        \emph{because} of the abstraction gap.
    }
    \label{fig:graphical-behaviour}
\end{figure*}

\firstReview{A pictorial representation of the experiment in form of simulation snapshots is shown in \Cref{fig:graphical-behaviour}.
For the sake of conciseness,
we do not provide the complete \ac{MAS} code here,
but we highlight in \Cref{fig:code} the platform-agnostic specification and how it is bound with the
underlying execution platform.
We refer the interested reader to the companion artefact~\cite{zenodojakta}
available online\footnotemark for the complete codebase of the experiment.}

\begin{figure}
    \begin{minipage}[b]{.62\linewidth}
        \centering
\begin{lstlisting}[
            language=Kotlin,
            linewidth=\linewidth,
            basicstyle=\scriptsize\ttfamily,
            label={lst:logic},
    %        caption={Platform-agnostic code}
        ]
fun AgentScope.leader() {
  goals { achieve("move") }
  plans {
    +achieve("move") then {
      execute("circleMovementStep")
      execute("notifyAgent")
      achieve("move")
    }
  }
}

fun AgentScope.follower() {
  plans {
    val shape = "joinCircle"(C, R, N)
    +achieve(shape) then {
      execute("follow"(C, R, N))
    }
  }
}
\end{lstlisting}
    \end{minipage}\hfill
    \begin{minipage}[b]{.34\linewidth}
        \centering
\begin{lstlisting}[
            language=Kotlin,
            linewidth=\linewidth,
            basicstyle=\scriptsize\ttfamily,
%        caption={Entrypoint},
            label={lst:agents}
        ]
agent("leader") {
  // Platform code
  ...
  // Shared logic
  leader()
}

agent("follower") {
  // Platform code
  ...
  // Shared logic
  follower()
}
\end{lstlisting}
    \end{minipage}
    \caption{
        Code extracts from the companion artifact.
        The agent specification (left) is completely platform-agnostic and reusable,
        some glue code (right) wires the logics with the underlying execution platform.
    }
    \label{fig:code}
\end{figure}

\paragraph{\firstReview{Configuring the Agent Execution and Simulation Parameters}}

\firstReview{We investigate how the mapping granularity (cf. \Cref{ssec:granularity}) impacts the system's behaviour.
To do so, we use \ac{AMA} as baseline,
letting the entire \ac{MAS} run a full cycle every simulated second
    (Dirac Comb distribution with frequency $f=1Hz \rightarrow T=1s$, cf. \Cref{subsec:mapping}),
thus replicating the behaviour of most current agent-based simulation frameworks, which are time-driven.
We compare the baseline with \ac{ACLI} and \ac{ACLP},
for which, however, we model
each agent's control-loop frequency following a Weibull distribution with mean $f$ and deviation $f\cdot\tau$ (drift).
Intuitively, this means that most loops will be scheduled around the mean frequency $f$, but some loops will be scheduled earlier or later, thus introducing a drift $\tau$ in the agents' execution.}

\firstReview{Additionally, for the \ac{ACLP} granularity we model deliberation and action delays,
associating each phase with an exponential distribution with rate $\lambda = f$:
faster agents (larger $f$ values) have less delay.
This is needed to emulate a real-world concurrent deployment, in which different phases of the control loop may take different time to complete and may interleave.}

\firstReview{Every experiment is repeated $100$ times with a different random seed, changing
the initial positions of the followers
and the distribution in time of the events for the \ac{ACLI} and \ac{ACLP}.
In each experiment, the leader follows a circular trajectory of radius $r_l=5m$
and is set to complete a full circle in $600s$.
We let the system execute for $1500$ simulated seconds.}

\subsubsection{\firstReview{Real-World System Deployment as a Concurrent System}}
\label{ssec:real-world-execution}

\firstReview{
    To demonstrate the seamless execution of the same \ac{MAS}
    specification in both simulated and real systems,
    we run the \jakta{} \ac{MAS} natively on a stand-alone desktop computer.
    The execution model in \jakta{} runs each agent in a separate thread,
    thereby emulating a concurrent deployment on independent \acp{UAV}.
}

\firstReview{
    In this setup,
    \acp{UAV} run at the maximum allowed frequency $f$,
    determined by the host machine's CPU capabilities.
    Any relative drift $\tau$ between different \acp{UAV}
    is naturally introduced by the operating system's scheduler.
    The duration of each phase of the control loop depends
    on the actual computation time of the agent program,
    which in turn is determined by the host machine's CPU performance.
}

\firstReview{
    Agents interact with a local instance of the environment
    implemented as a specialization of the generic \jakta{} \texttt{Environment}
    (cf.~\Cref{par:jakta}).
    The environment is responsible for emulating the \acp{UAV}' physical aspects,
    such as position and communication.
}

\firstReview{
    Concerning communication between \acp{UAV},
    agents exchange messages via the environment:
    each agent invokes the broadcast primitive provided by the environment,
    which delivers the message to all agents whose current
    positions lie within the sender's communication radius $r_c=5m$.
}

\firstReview{
    Concerning positioning,
    the environment stores \acp{UAV} positions as two-dimensional coordinates.
    When an agent attempts perceiving its own position,
    the environment returns the stored coordinates.
    When an agent attempts to move to a target destination,
    it sends a request to the environment,
    which in turns stores that desired destination for that agent.
    The actual movement is completed asynchronously via a co-routine which runs in background.
    This co-routine would periodically (as frequently as possible) update the positions of each \ac{UAV}
    along the straight line toward its target destination,
    by a distance non-greater than $1m$
    (so that $v_{\max}=1\nicefrac{m}{s}$).
}

\firstReview{
    This setup demonstrates the portability of the \ac{MAS}
    codebase to a non-simulated deployment:
    the same code used in simulation runs unchanged here.
    The experiment\footnotemark[\value{footnote}]
    shows the identical codebase executed in both simulated and real-world settings.
    %
    %
    The experiment was executed on a machine with an Intel i7-10750H CPU (2.60 GHz) and 32 GB RAM\@.
}
\footnotetext{
    \url{https://github.com/anitvam/uav-circle-2024-jakta-alchemist}
}

\subsection{Results}
\label{ssec:results}

In this section we present and discuss the results of our experiments,
focusing on the impact of the simulation granularity on the system behaviour.
We discuss the results obtained when simulating the system first,
then we present the results obtained from the concurrent system execution.

\firstReview{
We evaluate the system performance by measuring the deviation from the ideal positions.
More precisely,
we use an oracle to determine the ideal position of each follower based on the current leader's position,
then,
for each follower $u$,
we measure the distance error $d_u$ between the ideal and actual position of $u$.
%
Our error metric is the overall
squared distance error:
}
$\sum_{u \in U} d_u^2$.

Results confirm the expectations that
when simulating coarse-grained granularity levels
(namely, \ac{AMA} and \ac{ACLI})
the agents are able to maintain the formation,
while the fine-grained \ac{ACLP} granularity level reveals
that this result is an artefact of
the over-simplification of the \ac{BDI} programs.
On a real \ac{UAV}-programming workflow,
this would have revealed a flaw in the agent program,
or highlighted the need for higher responsiveness
    (reasoning cycle frequency)
to react to the dynamics of the environment.

\subsubsection{\firstReview{Analysis of the \alchemist{} Simulated System Deployment}}

\begin{figure*}
    \centering
    \includegraphics[width=\linewidth]{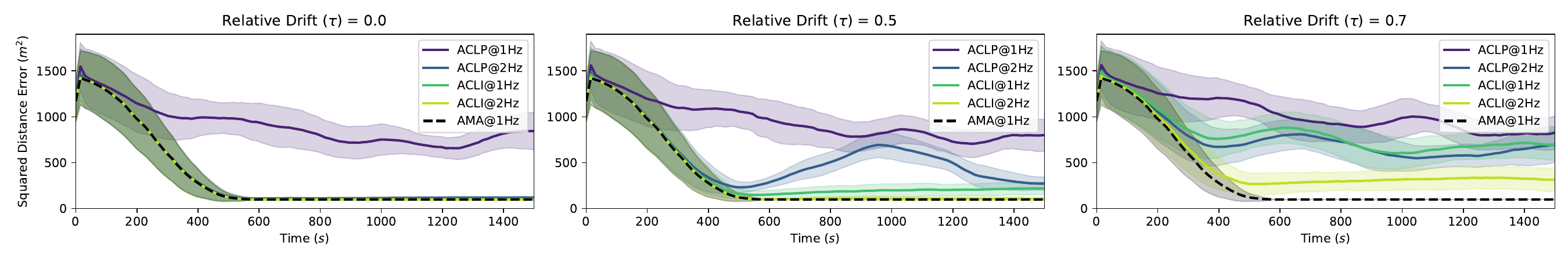}
    \caption{
        Error with time for different granularities, with \ac{ACLI} and \ac{ACLP} running at $f=1Hz$ and $f=2Hz$.
        Different charts show different values of relative drift $\tau$.
        Coloured shadows represent $\pm 1\sigma$.
    }
    \label{subfig:err_time}
\end{figure*}
\begin{figure*}
    \centering
    \includegraphics[width=\linewidth]{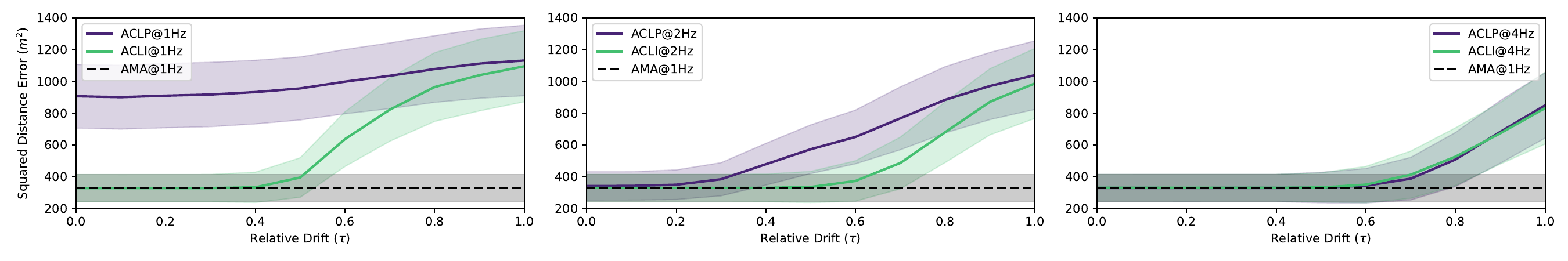}
    \caption{
        Mean squared distance error with relative drift ($\tau$), measured for different frequencies for \ac{ACLI} and \ac{ACLP}.
        Coloured shadows represent $\pm 1\sigma$.
    }
    \label{subfig:err_drift}
\end{figure*}

\Cref{subfig:err_time} shows the evolution in time results for $\tau \in \{0, 0.5, 0.7\}$.
With \ac{AMA} granularity,
after a transient phase in which the leader is still contacting the followers,
the error stabilises to a very low level,
due to the natural delay between the commands issued by the agent execution
and their realisation by the \ac{UAV}.
When $\tau=0$, \ac{ACLI} shows the same behaviour for both $f=1Hz$ and $f=2Hz$:
the system can apparently cope with the problem at hand.
%
\firstReview{As expected, given that the simplistic agent logic we implemented relies on implicit synchronisation,}
when using the most fine-grained model (\ac{ACLP}),
which is capable of modelling delays between one agent phase and another,
\firstReview{the simulation reveals that the error compared to the ideal positions is much larger.}
\firstReview{Indeed,}
the system seems to be able to cope with the problem at hand only when the agent runs at $2Hz$:
we have thus evidence that there are cases in which capturing a finer grained model
can provide insights
on the system's behaviour
that are not visible with a coarser granularity,
potentially imposing additional design or deployment constraints
(in this case, we must execute at at least $2Hz$ if we want the system to be robust to delays).

\firstReview{
    This result is important:
    since the \acp{UAV} in the target deployment would run independently,
    the \ac{ACLP} granularity is the most realistic one.
    If the developers had simulated the system with \ac{ACLI}
    (or, worse, \ac{AMA}) granularity,
    the system would have failed when deployed,
    despite positive feedback from the simulation.
}

If we introduce relative drift in the agent's execution frequency
\firstReview{-- thus improving the realism of the simulation --},
we can observe errors even with the \ac{ACLI} model.
With $\tau=0.5$, both \ac{ACLI} and \ac{ACLP} show that the system can work reasonably well, but only with $f=2Hz$.
Raising $\tau$ further shows that capturing the agent phases is relevant,
as the performance of \ac{ACLI} and \ac{ACLP} differs also for $f=2Hz$.

The impact of $\tau$ on the system is better evidenced in \Cref{subfig:err_drift},
where we sample multiple values of $\tau$ for different $f$.
We
\firstReview{demonstrate} that the \ac{ACLI} model provides unreliable results at lower frequencies,
while it is comparable to \ac{ACLP} as the frequency increases.
When the system is challenged, and the frequency is barely insufficient to cope with the problem,
it tends to provide overly optimistic results,
thus making the system appear as working as intended when, with better detail, we can observe that it is not.
Designers can use this information to decide whether their software needs to be amended,
or they may clearly state that it is required for the target \ac{UAV} to be capable to run the agent software
at a minimum frequency.

\subsubsection{\firstReview{Analysis of the Real-World Concurrent System Deployment}}
\label{sub-ssec:concurrent-results}
\begin{figure}
    \centering
    \includegraphics[width=\linewidth]{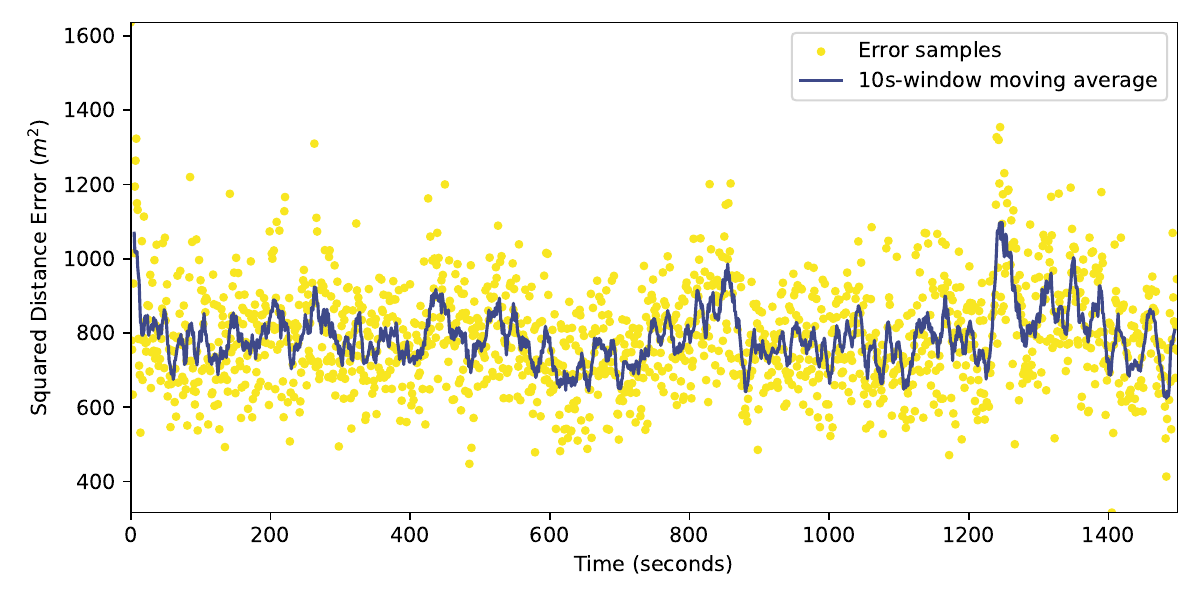}
    \caption{
        Error over time for the execution of the \ac{BDI} \ac{MAS} on a concurrent system.
        Yellow line represents the raw error measurements,
        gathered once every second.
        Blue line is the rolling average over a window of $10s$.
    }
    \label{fig:real-world-error}
\end{figure}

\firstReview{
    The execution of the same \ac{MAS} logic on a concurrent system
    produced results comparable to those obtained in simulation at the \ac{ACLP} granularity,
    demonstrating that simulation granularity affects the system's evaluation.
    The error metric for the concurrent-system execution was measured once per second,
    and the outcome is shown in \Cref{fig:real-world-error}.
}

\firstReview{
    In line with expectations,
    the error follows the \ac{ACLP} simulation results,
    confirming that coarser-grained event granularities in simulation
    may obscure subtle concurrency-dependent effects.
    This is confirmed by results shown in \Cref{fig:real-world-error},
    where the error does not stabilise or decrease towards zero,
    which is instead the behaviour we would expect from a correct system.
}

\subsubsection{Discussion}

With this experiment, we demonstrated that
the execution of a \ac{BDI} \ac{MAS} in a \ac{DES} is \emph{feasible}
and can be realised at different granularity.
We also show that the \ac{BDI} code
can be ported \emph{transparently} into the simulation environment,
as it does not directly refer any simulator- or platform-specific entity.
\Cref{fig:code} shows that the code of \ac{BDI} specification (left) is completely agnostic of the final destination of the agent
and programmed using pure \jakta{} syntax,
that can run both in a simulation and in a real-world scenario,
provided that the necessary interactions with the environment
(e.g., the action \texttt{circleMovementStep}) are appropriately modelled in both settings.
Finally, we show that the impact of the chosen granularity on the \emph{reality gap} is a relevant issue
and that executing a \ac{MAS} in modes that induce \emph{implicit synchronisation}
(\ac{AMA}, and, to a lesser extent, \ac{ACLI})
can produce unreliable results,
especially concerning the system's performance.

The example highlights
how simulating the \ac{MAS} behaviour before deployment
can be a valuable tool for software engineers,
allowing to identify potential issues.
In this case, the simulation provides insights
on the relative frequency agents need to run to cope with the problem.
Having observed that the system may fail under certain conditions,
the developers of the agent specification may decide to introduce additional mechanisms
to improve the system's reliability.

\section{Threats to validity}
\label{sec:limitations}


Here we briefly comment on a few critical aspects of our study design,
which may appear as threats to the general validity of the proposed approach.
We motivate design choices, and discuss how specific threats have been mitigated,
or why they have been accepted for the current work
with the plan of addressing them in future research.

\paragraph{Specificity of technological choices}

Our prototype is tailored on \jakta{} and \alchemist{},
which are technologies which have been designed and developed by some authors of this paper.
While this can be seen as a threat to the generalization of our results,
we argue that addressing our research questions
is also a technological task,
which requires technological commitments.
While we try to keep the conceptual discussion about the mapping
(\Cref{sec:bdi-over-des})
as general as possible,
we preferred to rely on technologies we know well when implementing the mapping.

Furthermore,
the technological affinities between \jakta{} and \alchemist{},
were key enablers for the development of this research line,
and this is yet another reason why we chose them.

\paragraph{Comparisons with other methods}

We deliberately avoid comparisons with other approaches for \ac{BDI} simulation\firstReview{.}
The reason behind that is that,
as discussed in \Cref{sec:related-works},
we are not aware of any prior attempt to run \ac{BDI} \acp{MAS} on \acp{DES}
with the same level of granularity we propose
and comparisons with other approaches would be unfair.
As the existing frameworks for \ac{BDI} simulation are mainly based on \ac{DTS},
we detailed conceptual differences with \ac{DTS} frameworks.
Yet,
we could not see the benefits of benchmarking our approach against other approaches based on \ac{DTS}.
In particular,
we expect \ac{DTS} would serve as a good engineering tool too,
despite being less adequate to study the effect of simulation granularity on the \ac{BDI} engineering process.

\paragraph{Performance analysis}

We do not study our solution from a computational complexity perspective,
and we do not investigate the performance of the simulation.

At the conceptual level,
our goal is to study the impact of simulations in the validation of \ac{BDI} systems,
and the role of \ac{DES} in this process.
We take an angle which is rooted in software engineering,
so we are more interested in studying what \ac{BDI} engineers can do with simulations,
rather than how fast the simulations run.
That said,
in our framework,
the computational cost of simulations is straightforward to estimate qualitatively:
fine-grained granularity levels will be more computationally expensive than coarse-grained ones,
as they require more events to be simulated.

Other practical considerations about performance are actually inherited to the technological choices we made.
\jakta{} is a lightweight \ac{BDI} framework,
still in its infancy,
but under active development,
and performance analysis are part of the roadmap.
\alchemist{} is a mature \ac{DES} framework,
which has been used in the past to simulate large-scale systems,
as proved by the many publications that rely on it for their experiments.
So,
in this work,
performance is not a concern,
and not even a controllable variable.

\paragraph{Effort required to design and run a testing scenario}

The byproduct of our feasibility study is a prototype tool
that may be interesting for the community of \ac{BDI} developers.
Despite the usability of the tool is not the focus of this paper,
we understand that it might be interesting to evaluate it
to understand how easy it is to run a \ac{BDI} simulation via \jakta{}+\alchemist{}.
We consider refining and evaluating the usability of the tool in future works,
but we can already anticipate that the effort required to design and run a testing scenario
using our approach involves:
\begin{inlinelist}
    \item\label{step:jakta-code} writing agent and, optionally, action specifications, in \jakta{};
    \item\label{step:environment-code} coding the simulated environment, in Kotlin;
    \item\label{step:alchemist-code} configuring simulations using Alchemist, via YAML;
    \item\label{step:textual-results} run the simulations and analyse simulation traces as text files.
\end{inlinelist}

Step \ref{step:jakta-code} involves writing Kotlin code using \jakta{}'s \ac{API}
(as \jakta{} technically consists of an internal \ac{DSL} for Kotlin~\cite{SNCS:JaktaJournal}).
Its difficulty depends on the scenario at hand, and on the goal of the programmers.
In principle, this step is analogous to writing \ac{BDI} agents in any other \ac{BDI} framework,
e.g. Jason~\cite{BordiniHW2007}.
Also, this step would be required in any case, even if the \ac{MAS} is going to be deployed directly with no simulation.

Step \ref{step:environment-code} is required if the \ac{BDI} system is going to be tested in a simulated environment.
This may require additional design and coding effort, if designers want their simulation to be high-fidelity.

Step \ref{step:alchemist-code} is required to configure the simulation.
This is as simple as creating a YAML file where the parameters of the simulation are set.

Finally,
step \ref{step:textual-results} is required to analyse the results of the simulation,
and this may require additional custom processing,
depending on the goals of the users.



\paragraph{Simplified scenario / Scalability}

The UAV scenario in \Cref{sec:evaluation} has not been tested on real \acp{UAV}.
\firstReview{The experiment is not meant to demonstrate the applicability of \ac{BDI} systems to real-world
\ac{UAV} coordination problems, which may require additional considerations beyond the scope of this work.}
\firstReview{However, i}t does demonstrate how simulation positively impacts \ac{BDI} system validation
and how execution granularity affects fidelity.

\firstReview{Additionally, the demonstration of execution of the same \ac{MAS} codebase as
a concurrent system on the host machine through the \jakta{} built-in threading model
sufficiently shows the portability of the codebase to real-world deployments.}
Indeed,
most of the technical effort in this work
has been to make \jakta{} agents run in a simulation environment \emph{too},
\firstReview{as \jakta{} is designed for programming \ac{BDI} agents to be deployed in the real world
as multithreaded applications.}
%

%
\firstReview{The choice of the \acp{UAV} scenario was driven by multiple factors.}
\firstReview{W}e decided to focus on an example
in which the \ac{BDI} system is complex enough to exhibit different dynamics at different event granularities
so that the reader could qualitatively appreciate the impact of the granularity on the system's behaviour.
\firstReview{Moreover},
we wanted to emulate a real scenario where the effort of validating the \ac{MAS} via simulation
is justified by the cost of deploying the real system,
and the \ac{UAV} scenario perfectly fits this purpose.

\firstReview{Regarding scalability, the chosen scenario involves a limited number of agents
 (17 in total: 1 leader and 16 followers)
 to keep the simulation manageable and interpretable for demonstration purposes.}
This should not be interpreted as a limitation to the scalability of our approach:
as we rely on the \alchemist{} simulator,
our approach may easily scale to large-scale systems composed by thousands of agents (or more).

\section{Conclusion and Future Work}\label{sec:conclusion}

In this paper,
we analysed the role of simulation in the engineering of distributed \ac{BDI} \acp{MAS},
discussing the importance of simulating the same agent specification that will be deployed in the real world,
and showing the practical implications of mapping \ac{BDI} agents on \acp{DES} at different granularity.
We substantiate our claims by producing an open-source prototype that maps a \ac{BDI} framework onto a \ac{DES} simulator,
using which we can execute an agent specification in a simulated environment with no changes to the agents' code.
We leverage the prototype to show the impact of the granularity on the simulation's reality gap.

The paper shares two main insights for software engineers designing \acp{MAS}.
\begin{enumerate}[wide,labelwidth=!,labelindent=0pt]
    \item Simulation is a crucial tool in the development of \ac{BDI} \acp{MAS}, and \ac{BDI} framework designers should consider it upfront.
    Designing a \ac{BDI} engine that does not provide appropriate \acp{API} to select fine-enough granularities,
    in fact,
    may prevent the system's behaviour and performance from being correctly assessed ahead of deployment.
    \item The \ac{MAS} under development must be validated using the same code used that will be deployed,
    reducing the time required to test the system
    and easing software maintenance through simulation-based regression testing.
\end{enumerate}

\subsection{\firstReview{Achievement of the work objectives}}\label{subsec:answering-the-research-questions}

\firstReview{
We here comment regarding the sub-objectives listed in \Cref{sec:introduction}.
}

\paragraph{\ref{rq:methods}: \Paste{RQ1text}}

\firstReview{At the design level,}
two simulation paradigms can be used to develop and test \ac{BDI} agents:
\emph{time-driven} simulation and \emph{event-driven} simulation,
the latter commonly implemented as \ac{DES}.
Time-driven simulation advances in fixed time steps, updating all agents at each step,
which simplifies implementation and parallel execution
(at the price of reducing reproducibility)
but can introduce artificial synchronisation
and unnecessary computations when no relevant events occur.
In contrast,
event-driven simulation advances only when meaningful events occur,
allowing finer-grained control of agent execution and reducing computational overhead.
\ac{DES} is particularly well-suited for \ac{BDI} agents,
as their reasoning cycle is inherently event-driven,
making it a natural fit for mapping the agent’s execution flow into a simulation model.
However,
this approach introduces trade-offs:
while \ac{DES} improves fidelity by accurately modelling asynchronous agent interactions,
it can be more complex to implement and requires careful event scheduling
to ensure reproducibility and scalability.
Our work demonstrates that \ac{DES} provides an effective balance between
\firstReview{implementation complexity} and simulation accuracy,
making it a promising choice for testing \ac{BDI} systems before deployment.

\firstReview{
At the technical level,
simulating \ac{BDI} agents has been supported in the literature by
adding simulation-related features to existing \ac{BDI} interpreters,
by adding \ac{BDI}-related features to existing simulation platforms,
or by synchronising two separate tools
(one for \ac{BDI} and one for simulation).
Seeking for some solution which does not require to rewrite the agent specification
upon switching from simulation to real-world deployment (or vice versa)
while retaining control over the fidelity of the simulation,
we identified another viable approach where a new sort of \ac{BDI} interpreter is natively designed
to allow for both in-silico and real-world execution%
---and this is the approach proposed by our work.
}

\paragraph{
    \ref{rq:abstract-spec}:
    \Paste{RQ2text}
}

A \ac{BDI} framework must support modular execution models,
abstracting platform-specific functionalities and encapsulating them into a deployment-agnostic \ac{API},
that allows external scheduling of agents' reasoning cycles.
The framework should maintain a clean separation between agent logic and execution infrastructure,
allowing the same specification to run both in simulation and in the real-world target environment
without modifications.
Implementations should interpret the same agent logic differently.
Core to the \emph{simulated} deployment is
the ability to execute \ac{BDI} agents deterministically,
to ensure reproducibility.
The same is not true in \emph{real-world deployments} where concurrency and asynchrony are inherent,
and sometimes even desirable and exploitable.
Therefore,
the deployment-agnostic \ac{API} should \emph{virtualise} functionalities covering all potential sources of non-determinism,
such as random number generation and external events.


\paragraph{
    \ref{rq:des}:
    \Paste{RQ3text}
}

\ac{BDI} agent execution can be mapped onto \ac{DES} by treating key stages of the reasoning cycle as discrete events,
scheduled within the simulator.
We identify four mapping granularities, each with distinct implications:
\begin{itemize}
    \item \textbf{\ac{AMA}} (\Cref{subsubsec:ama}):
    The most coarse-grained approach, synchronising all agents at each \ac{DES} event,
    simplifying implementation but discarding interleaving effects.
    \item \textbf{\ac{ACLI}} (\Cref{subsubsec:acli}):
    A finer-grained model where each agent executes its full control loop independently,
    allowing action interleaving but not asynchronous phase execution.
    \item \textbf{\ac{ACLP}} (\Cref{subsubsec:aclp}):
    Each atomic phase (sense, deliberate, act) is a distinct \ac{DES} event,
    preserving concurrency and accurately modelling distributed agent behaviour.
    \item \textbf{\ac{ABE}} (\Cref{subsubsec:abe}):
    The finest-grained approach, mapping each \ac{BDI} event to a \ac{DES} event,
    useful for debugging but behaviourally equivalent to \ac{ACLP}.
\end{itemize}
Our findings demonstrate that \ac{BDI} execution can be faithfully represented in \ac{DES},
with granularity selection impacting fidelity, performance, and complexity.
\ac{ACLP} ensures the highest accuracy but at a computational cost,
whereas \ac{AMA} and \ac{ACLI} simplify execution but risk oversimplifying agent interactions.
The appropriate mapping depends on the testing objectives and the trade-offs
between efficiency and behavioural fidelity.

\paragraph{
    \ref{rq:sim-integration}:
    \Paste{RQ4text}
}

Our integration of \jakta{} and \alchemist{} demonstrates that a \ac{BDI} framework can be embedded
within a \ac{DES} engine by treating agent execution as a series of simulation-driven events.
By leveraging modularity in both technologies, we enable \ac{BDI} agents to operate in a simulated
environment without requiring code modifications.
The environment is modeled within the simulator, preserving interactions with agents while ensuring portability.
The proposed approach ensures that the same \ac{BDI} specification can be tested in simulation
and seamlessly deployed in the real-world environment
(e.g., on distributed computing devices)
reducing development overhead and minimizing the reality gap.

%


\subsection{Future Research Directions}\label{subsec:future-research-directions}

This work opens multiple research directions that we plan to investigate in the future.

\paragraph{Debugging and Automated Testing}
Complex software systems are better inspected if the abstractions
used to design a system are exposed first-class during inspection,
and if the inspection tool supports the injection of dynamic events from the environment.
Unfortunately,
conventional debug tools fall short in both aspects,
as they expose low-level abstractions when debugging \ac{BDI} programs,
they allow solely for the execution of the program in the local device
(often not representative of the real-world environment),
and do not capture/inject external events easily.
Simulation can solve the dynamicity issue by
emulating multiple interacting devices at once:
it can thus represent the basis towards a debugging tool for (distributed) BDI MASs.
However, the first element remains open to investigation, raising multiple questions:
\begin{inlinelist}
    \item how do breakpoint work when the \ac{BDI} system is being simulated?
    \item How does such behaviour is compared to stand-alone execution?
    \item Which abstractions should be exposed by the debugger?
\end{inlinelist}
Moreover,
since simulation can be a valid option for testing \ac{BDI} \ac{MAS},
we believe it is useful to integrate simulation in automated testing pipelines,
to verify \ac{MAS} compliance with the expected outcomes.
Along this line,
the very concept of ``successful simulation/test''
deserves further investigation, especially in the context of stochastic systems.

\paragraph{Static Checking of the Simulated Environment}
For the simulation to run successfully,
the environment must provide the necessary \ac{API} for the agents to run.
If some of them is missing the error will be raised at runtime,
potentially even after many successful runs due to the stochastic nature of the simulation.
This failure can be prevented with statically verifying the simulation setup,
making sure that all the actions required by the \ac{BDI} \ac{MAS} are available.
We plan to explore how to implement these checks
via static analysis or through the type system,
to prevent failures at runtime which could make the verification process slower and less reliable.

\paragraph{Adaptive Fidelity in Simulation-Based Testing}

Investigating techniques to dynamically adjust the granularity of \ac{DES} execution
could improve efficiency without compromising fidelity.
Adaptive models could vary the simulation resolution based on runtime conditions,
prioritising high fidelity only in critical interaction phases.

\paragraph{Hybrid simulation}

Exploring hybrid approaches that combine simulated and real-world execution
could provide a smoother transition from testing to deployment.
Methods such as \ac{HiL} simulation~\cite{DBLP:journals/tec/KoosMD13}
may help assess the impact of physical constraints on \ac{BDI} agent execution,
especially when the deployment targets are robots.

\paragraph{Standardisation and Interoperability}

Developing standardised interfaces for integrating \ac{BDI} frameworks with simulation engines
could enhance interoperability and facilitate adoption in different application domains.
Defining common APIs and data exchange formats would improve reusability across tools.

\paragraph{Practical application to robot fleets}
The construction of a demonstrator with the proposed approach
applied to real-world small-scale \acp{UAV} or \acp{UGV} fleets
could validate the practical applicability of the approach.
However, this requires addressing challenges related to porting the prototype
to low-power devices, network communication, and real-time constraints.
Although we have a preliminary demonstrator with \acp{UGV}~\cite{DBLP:conf/coordination/AguzziBBCCDFPV25},
the current prototype needs further refinement to be effectively deployed on real-world robot fleets.



\backmatter





\bmhead{Acknowledgements}

This work has been partially supported by:
``WOOD4.0 - Woodworking Machines for Industry 4.0'', (CUP E69J22007520009) Emilia-Romagna regional project, call 2022, art. 6 L.R. N. 14/2014;
and by the projects funded by the European Commission under the NextGenerationEU programme:
PNRR – M4C2 – Investment 1.3, Partenariato Esteso PE00000013 – ``FAIR—Future Artificial Intelligence Research'' – Spoke 8 ``Pervasive AI''
and
PNRR M4C2, Investment 3.3 (DM 352/2022) in collaboration with AUSL Romagna - CUP J33C22001400009 ``Digital Twins Ecosystems for the clinical, strategic and process governance in Healthcare''.

\bibliography{bibliography}

@article{FigueiredoPrudencio2024,
  title = {A Survey on Offline Reinforcement Learning: Taxonomy,  Review,  and Open Problems},
  volume = {35},
  ISSN = {2162-2388},
  _url = {http://dx.doi.org/10.1109/TNNLS.2023.3250269},
  DOI = {10.1109/tnnls.2023.3250269},
  number = {8},
  journal = {IEEE Transactions on Neural Networks and Learning Systems},
  publisher = {Institute of Electrical and Electronics Engineers (IEEE)},
  author = {Figueiredo Prudencio,  Rafael and Maximo,  Marcos R. O. A. and Colombini,  Esther Luna},
  year = {2024},
  month = aug,
  pages = {10237–10257}
}

@article{Watkins1992,
  title = {Q-learning},
  volume = {8},
  ISSN = {1573-0565},
  _url = {http://dx.doi.org/10.1007/BF00992698},
  DOI = {10.1007/bf00992698},
  number = {3–4},
  journal = {Machine Learning},
  publisher = {Springer Science and Business Media LLC},
  author = {Watkins,  Christopher J. C. H. and Dayan,  Peter},
  year = {1992},
  month = may,
  pages = {279–292}
}

@book{Wooldridge00,
  author    = {Michael J. Wooldridge},
  title     = {Reasoning about rational agents},
  series    = {Intelligent robots and autonomous agents},
  publisher = {{MIT} Press},
  year      = {2000},
  isbn      = {978-0-262-23213-5},
  bibsource = {dblp computer science bibliography, https://dblp.org},
  address   = {Cambridge, MA}
}

@misc{zenodojakta,
  doi       = {10.5281/ZENODO.13921206},
  _url      = {https://zenodo.org/doi/10.5281/zenodo.13921206},
  author    = {Martina Baiardi and Danilo Pianini},
  title     = {anitvam/uav-circle-2024-jakta-alchemist: 1.0.4},
  publisher = {Zenodo},
  year      = {2024},
  copyright = {Creative Commons Attribution 4.0 International}
}

@article{CalegariCMO21,
  author    = {Roberta Calegari and
               Giovanni Ciatto and
               Viviana Mascardi and
               Andrea Omicini},
  title     = {Logic-based technologies for multi-agent systems: a systematic literature
               review},
  journal   = {Auton. Agents Multi Agent Syst.},
  volume    = {35},
  number    = {1},
  pages     = {1},
  year      = {2021},
  _url      = {https://doi.org/10.1007/s10458-020-09478-3},
  doi       = {10.1007/S10458-020-09478-3},
  timestamp = {Fri, 14 May 2021 08:32:14 +0200},
  biburl    = {https://dblp.org/rec/journals/aamas/CalegariCMO21.bib},
  bibsource = {dblp computer science bibliography, https://dblp.org}
}

@article{IngrandGR1992,
  author  = {Ingrand, F.F. and Georgeff, M.P. and Rao, A.S.},
  journal = {IEEE Expert},
  title   = {An architecture for real-time reasoning and system control},
  year    = {1992},
  volume  = {7},
  number  = {6},
  pages   = {34-44},
  doi     = {10.1109/64.180407}
}

@article{DInvernoLGKW04,
  author    = {Mark d'Inverno and
               Michael Luck and
               Michael P. Georgeff and
               David Kinny and
               Michael J. Wooldridge},
  title     = {The dMARS Architecture: {A} Specification of the Distributed Multi-Agent
               Reasoning System},
  journal   = {Auton. Agents Multi Agent Syst.},
  volume    = {9},
  number    = {1-2},
  pages     = {5--53},
  year      = {2004},
  _url      = {https://doi.org/10.1023/B:AGNT.0000019688.11109.19},
  doi       = {10.1023/B:AGNT.0000019688.11109.19},
  timestamp = {Fri, 13 Mar 2020 10:56:03 +0100},
  biburl    = {https://dblp.org/rec/journals/aamas/DInvernoLGKW04.bib},
  bibsource = {dblp computer science bibliography, https://dblp.org}
}

@article{Xie2017,
  author     = {Jing Xie and Chen-Ching Liu},
  title      = {Multi-agent systems and their applications},
  journal    = {Journal of International Council on Electrical Engineering},
  volume     = {7},
  number     = {1},
  pages      = {188--197},
  year       = {2017},
  _publisher = {Taylor \& Francis},
  doi        = {10.1080/22348972.2017.1348890}
}

@article{CroattiMRGAA19,
  author    = {Angelo Croatti and
               Sara Montagna and
               Alessandro Ricci and
               Emiliano Gamberini and
               Vittorio Albarello and
               Vanni Agnoletti},
  title     = {{BDI} personal medical assistant agents: The case of trauma tracking
               and alerting},
  journal   = {Artif. Intell. Medicine},
  volume    = {96},
  pages     = {187--197},
  year      = {2019},
  _url      = {https://doi.org/10.1016/j.artmed.2018.12.002},
  doi       = {10.1016/J.ARTMED.2018.12.002},
  timestamp = {Sat, 30 Sep 2023 10:01:55 +0200},
  biburl    = {https://dblp.org/rec/journals/artmed/CroattiMRGAA19.bib},
  bibsource = {dblp computer science bibliography, https://dblp.org}
}

@inproceedings{IssicabaRPB17,
  author    = {Diego Issicaba and
               Mauro Augusto da Rosa and
               Alexander M. Prostejovsky and
               Henrik W. Bindner},
  title     = {Experimental validation of {BDI} agents for distributed control of electric power grids},
  booktitle = {{IEEE} {PES} Innovative Smart Grid
               Technologies Conference Europe (ISGT-Europe)},
  pages     = {1--6},
  publisher = {{IEEE}},
  address   = {Torino, Italy},
  year      = {2017},
  _url      = {https://doi.org/10.1109/ISGTEurope.2017.8260273},
  doi       = {10.1109/ISGTEUROPE.2017.8260273},
  timestamp = {Wed, 16 Oct 2019 14:14:48 +0200},
  biburl    = {https://dblp.org/rec/conf/isgteurope/IssicabaRPB17.bib},
  bibsource = {dblp computer science bibliography, https://dblp.org}
}

@inproceedings{SilvestreLDBHB23,
  author     = {Iago de Oliveira Silvestre and
                Bruno de Lima and
                Pedro Henrique Dias and
                Leandro Buss Becker and
                Jomi Fred H{\"{u}}bner and
                Maiquel de Brito},
  _editor    = {Philippe Mathieu and
                Frank Dignum and
                Paulo Novais and
                Fernando de la Prieta},
  title      = {{UAV} Swarm Control and Coordination Using Jason {BDI} Agents on Top
                of {ROS}},
  booktitle  = {{PAAMS} 2023,
                Proceedings},
  _booktitle = {Advances in Practical Applications of Agents, Multi-Agent Systems,
                and Cognitive Mimetics. The {PAAMS} Collection - 21st International
                Conference, {PAAMS} 2023, Guimar{\~{a}}es, Portugal, July 12-14, 2023,
                Proceedings},
  _series    = {Lecture Notes in Computer Science},
  volume     = {13955},
  pages      = {225--236},
  _publisher = {Springer},
  year       = {2023},
  _url       = {https://doi.org/10.1007/978-3-031-37616-0_19},
  doi        = {10.1007/978-3-031-37616-0_19},
  timestamp  = {Fri, 14 Jul 2023 22:01:37 +0200},
  biburl     = {https://dblp.org/rec/conf/paams/SilvestreLDBHB23.bib},
  bibsource  = {dblp computer science bibliography, https://dblp.org}
}

@article{BandiniMV09,
  author    = {Stefania Bandini and
               Sara Manzoni and
               Giuseppe Vizzari},
  title     = {Agent Based Modeling and Simulation: An Informatics Perspective},
  journal   = {J. Artif. Soc. Soc. Simul.},
  volume    = {12},
  number    = {4},
  year      = {2009},
  url       = {http://jasss.soc.surrey.ac.uk/12/4/4.html},
  bibsource = {dblp computer science bibliography, https://dblp.org}
}

@inproceedings{DrogoulVM02,
  author     = {Alexis Drogoul and
                Diane Vanbergue and
                Thomas Meurisse},
  _editor    = {Jaime Sim{\~{a}}o Sichman and
                Fran{\c{c}}ois Bousquet and
                Paul Davidsson},
  title      = {Multi-agent Based Simulation: Where Are the Agents?},
  booktitle  = {{MABS} 2002},
  _booktitle = {Multi-Agent-Based Simulation, Third International Workshop, {MABS}
                2002, Bologna, Italy, July 15-16, 2002, Revised Papers},
  _series    = {Lecture Notes in Computer Science},
  volume     = {2581},
  pages      = {1--15},
  _publisher = {Springer},
  year       = {2002},
  _url       = {https://doi.org/10.1007/3-540-36483-8_1},
  doi        = {10.1007/3-540-36483-8_1},
  timestamp  = {Wed, 25 Sep 2019 18:15:39 +0200},
  biburl     = {https://dblp.org/rec/conf/mabs/DrogoulVM02.bib},
  bibsource  = {dblp computer science bibliography, https://dblp.org}
}

@article{Sklar07,
  author    = {Elizabeth Sklar},
  title     = {NetLogo, a Multi-agent Simulation Environment},
  journal   = {Artif. Life},
  volume    = {13},
  number    = {3},
  pages     = {303--311},
  year      = {2007},
  _url      = {https://doi.org/10.1162/artl.2007.13.3.303},
  doi       = {10.1162/ARTL.2007.13.3.303},
  timestamp = {Fri, 13 Mar 2020 10:59:42 +0100},
  biburl    = {https://dblp.org/rec/journals/alife/Sklar07.bib},
  bibsource = {dblp computer science bibliography, https://dblp.org}
}

@article{LukeCPSB05,
  author    = {Sean Luke and
               Claudio Cioffi{-}Revilla and
               Liviu Panait and
               Keith Sullivan and
               Gabriel Catalin Balan},
  title     = {{MASON:} {A} Multiagent Simulation Environment},
  journal   = {Simul.},
  volume    = {81},
  number    = {7},
  pages     = {517--527},
  year      = {2005},
  _url      = {https://doi.org/10.1177/0037549705058073},
  doi       = {10.1177/0037549705058073},
  timestamp = {Mon, 08 Jun 2020 22:17:39 +0200},
  biburl    = {https://dblp.org/rec/journals/simulation/LukeCPSB05.bib},
  bibsource = {dblp computer science bibliography, https://dblp.org}
}

@article{NorthCOTMBS13,
  author    = {Michael J. North and
               Nicholson T. Collier and
               Jonathan Ozik and
               Eric R. Tatara and
               Charles M. Macal and
               Mark J. Bragen and
               Pam Sydelko},
  title     = {Complex adaptive systems modeling with Repast Simphony},
  journal   = {Complex Adapt. Syst. Model.},
  volume    = {1},
  pages     = {3},
  year      = {2013},
  _url      = {https://doi.org/10.1186/2194-3206-1-3},
  doi       = {10.1186/2194-3206-1-3},
  timestamp = {Thu, 20 Aug 2020 22:49:22 +0200},
  biburl    = {https://dblp.org/rec/journals/casm/NorthCOTMBS13.bib},
  bibsource = {dblp computer science bibliography, https://dblp.org}
}

@inproceedings{kehoe2016robust,
  title     = {Robust reproducibility of agent based models},
  author    = {Kehoe, Joseph},
  booktitle = {The European Simulation and Modelling Conference. Inderscience},
  year      = {2016}
}

@book{BratmanEtAl1987,
  title     = {Intention, plans, and practical reason},
  author    = {Bratman, Michael and others},
  volume    = {10},
  year      = {1987},
  publisher = {Harvard University Press},
  address   = {Cambridge, MA}
}

@article{AdamGaudou2016,
  title     = {BDI agents in social simulations: a survey},
  author    = {Adam, Carole and Gaudou, Benoit},
  journal   = {The Knowledge Engineering Review},
  volume    = {31},
  number    = {3},
  pages     = {207--238},
  year      = {2016},
  publisher = {Cambridge University Press}
}

@article{NunesSF17,
  author    = {Ingrid Nunes and
               Frederico Schardong and
               Alberto E. Schaeffer Filho},
  title     = {BDI2DoS: An application using collaborating {BDI} agents to combat
               DDoS attacks},
  journal   = {J. Netw. Comput. Appl.},
  volume    = {84},
  pages     = {14--24},
  year      = {2017},
  _url      = {https://doi.org/10.1016/j.jnca.2017.01.035},
  doi       = {10.1016/J.JNCA.2017.01.035},
  timestamp = {Thu, 14 Oct 2021 09:11:23 +0200},
  biburl    = {https://dblp.org/rec/journals/jnca/NunesSF17.bib},
  bibsource = {dblp computer science bibliography, https://dblp.org}
}

@inproceedings{MontagnaSAC2012,
  author     = {Sara Montagna and
                Danilo Pianini and
                Mirko Viroli},
  title      = {A model for drosophila melanogaster development from a single cell
                to stripe pattern formation},
  _booktitle = {Proceedings of the {ACM} Symposium on Applied Computing, {SAC} 2012,
                Riva, Trento, Italy, March 26-30, 2012},
  booktitle  = {Proceedings of {ACM} {SAC} 2012},
  pages      = {1406--1412},
  year       = {2012},
  _url       = {http://doi.acm.org/10.1145/2245276.2231999},
  doi        = {10.1145/2245276.2231999},
  timestamp  = {Fri, 08 Jun 2012 19:37:23 +0200},
  biburl     = {http://dblp.uni-trier.de/rec/bib/conf/sac/MontagnaPV12},
  bibsource  = {dblp computer science bibliography, http://dblp.org}
}

@article{PianiniJOS2013,
  author    = {Danilo Pianini and
               Sara Montagna and
               Mirko Viroli},
  title     = {Chemical-oriented simulation of computational systems with {ALCHEMIST}},
  journal   = {J. Simulation},
  volume    = {7},
  number    = {3},
  pages     = {202--215},
  year      = {2013},
  _url      = {http://dx.doi.org/10.1057/jos.2012.27},
  doi       = {10.1057/jos.2012.27},
  timestamp = {Tue, 27 Aug 2013 14:29:37 +0200},
  biburl    = {http://dblp.uni-trier.de/rec/bib/journals/jos/PianiniMV13},
  bibsource = {dblp computer science bibliography, http://dblp.org}
}

@article{ZambonelliPMC2015,
  author    = {Franco Zambonelli and
               Andrea Omicini and
               Bernhard Anzengruber and
               Gabriella Castelli and
               De Angelis, Francesco L. and
               Di Marzo Serugendo, Giovanna and
               Dobson, Simon A. and
               Fernandez-Marquez, Jose Luis and
               Alois Ferscha and
               Marco Mamei and
               Stefano Mariani and
               Ambra Molesini and
               Sara Montagna and
               Jussi Nieminen and
               Danilo Pianini and
               Matteo Risoldi and
               Alberto Rosi and
               Graeme Stevenson and
               Mirko Viroli and
               Juan Ye},
  title     = {Developing pervasive multi-agent systems with nature-inspired coordination},
  journal   = {Pervasive and Mobile Computing},
  volume    = {17},
  pages     = {236--252},
  year      = {2015},
  _url      = {http://dx.doi.org/10.1016/j.pmcj.2014.12.002},
  doi       = {10.1016/j.pmcj.2014.12.002},
  timestamp = {Wed, 18 Mar 2015 14:32:55 +0100},
  biburl    = {http://dblp.uni-trier.de/rec/bib/journals/percom/ZambonelliOACAS15},
  bibsource = {dblp computer science bibliography, http://dblp.org}
}

@inproceedings{PianiniSAC2015,
  author     = {Danilo Pianini and
                Mirko Viroli and
                Jacob Beal},
  title      = {Protelis: practical aggregate programming},
  _booktitle = {Proceedings of the 30th Annual {ACM} Symposium on Applied Computing,
                Salamanca, Spain, April 13-17, 2015},
  booktitle  = {Proceedings of {ACM} {SAC}, 2015},
  pages      = {1846--1853},
  year       = {2015},
  _url       = {http://doi.acm.org/10.1145/2695664.2695913},
  doi        = {10.1145/2695664.2695913},
  timestamp  = {Tue, 21 Jul 2015 21:17:02 +0200},
  biburl     = {http://dblp.uni-trier.de/rec/bib/conf/sac/PianiniVB15},
  bibsource  = {dblp computer science bibliography, http://dblp.org}
}

@article{scafi,
  author    = {Roberto Casadei and
               Mirko Viroli and
               Gianluca Aguzzi and
               Danilo Pianini},
  title     = {ScaFi: {A} Scala {DSL} and Toolkit for Aggregate Programming},
  journal   = {SoftwareX},
  volume    = {20},
  pages     = {101248},
  year      = {2022},
  _url      = {https://doi.org/10.1016/j.softx.2022.101248},
  doi       = {10.1016/j.softx.2022.101248},
  timestamp = {Fri, 10 Feb 2023 23:34:31 +0100},
  biburl    = {https://dblp.org/rec/journals/softx/CasadeiVAP22.bib},
  bibsource = {dblp computer science bibliography, https://dblp.org}
}

@inproceedings{BaiardiBCP23,
  author     = {Martina Baiardi and
                Samuele Burattini and
                Giovanni Ciatto and
                Danilo Pianini},
  _editor    = {Vadim Malvone and
                Aniello Murano},
  title      = {JaKtA: {BDI} Agent-Oriented Programming in Pure Kotlin},
  booktitle  = {{EUMAS} 2023, Proceedings},
  _booktitle = {Multi-Agent Systems - 20th European Conference, {EUMAS} 2023, Naples,
                Italy, September 14-15, 2023, Proceedings},
  _series    = {Lecture Notes in Computer Science},
  volume     = {14282},
  pages      = {49--65},
  _publisher = {Springer},
  year       = {2023},
  _url       = {https://doi.org/10.1007/978-3-031-43264-4_4},
  doi        = {10.1007/978-3-031-43264-4_4},
  timestamp  = {Sat, 30 Sep 2023 09:40:42 +0200},
  biburl     = {https://dblp.org/rec/conf/eumas/BaiardiBCP23.bib},
  bibsource  = {dblp computer science bibliography, https://dblp.org}
}

@inproceedings{davoust_architecture_2020,
  address    = {Cham},
  title      = {An {Architecture} for {Integrating} {BDI} {Agents} with a {Simulation} {Environment}},
  isbn       = {978-3-030-51417-4},
  doi        = {10.1007/978-3-030-51417-4_4},
  language   = {en},
  booktitle  = {Engineering {Multi}-{Agent} {Systems}},
  _publisher = {Springer International Publishing},
  author     = {Davoust, Alan and Gavigan, Patrick and Ruiz-Martin, Cristina and Trabes, Guillermo and Esfandiari, Babak and Wainer, Gabriel and James, Jeremy},
  _editor    = {Dennis, Louise A. and Bordini, Rafael H. and Lespérance, Yves},
  year       = {2020},
  keywords   = {AgentSpeak Language (ASL), Architecture, Belief-Desire-Intention (BDI), Jason, Modeling and simulation},
  pages      = {67--84}
}

@article{singh_integrating_2016,
  title    = {Integrating {BDI} {Agents} with {Agent}-{Based} {Simulation} {Platforms}},
  volume   = {30},
  issn     = {1573-7454},
  _url     = {https://doi.org/10.1007/s10458-016-9332-x},
  doi      = {10.1007/s10458-016-9332-x},
  language = {en},
  number   = {6},
  urldate  = {2024-03-28},
  journal  = {Autonomous Agents and Multi-Agent Systems},
  author   = {Singh, Dhirendra and Padgham, Lin and Logan, Brian},
  month    = nov,
  year     = {2016},
  keywords = {Agent-based modelling, BDI, Integration, Simulation},
  pages    = {1050--1071}
}

@inproceedings{ricci_exploiting_2020,
  _series   = {Lecture {Notes} in {Computer} {Science}},
  title     = {Exploiting {Simulation} for {MAS} {Development} and {Execution} - {The} {JaCaMo}-{Sim} {Approach}},
  volume    = {12589},
  isbn      = {978-3-030-66533-3},
  _url      = {https://doi.org/10.1007/978-3-030-66534-0_3},
  doi       = {10.1007/978-3-030-66534-0_3},
  urldate   = {2024-03-28},
  booktitle = {Engineering {Multi}-{Agent} {Systems} - 8th {International} {Workshop}, {EMAS} 2020, {Revised} {Selected} {Papers}},
  publisher = {Springer},
  address   = {{Auckland}, {New} {Zealand}},
  month     = {May},
  author    = {Ricci, Alessandro and Croatti, Angelo and Bordini, Rafael H. and Hübner, Jomi Fred and Boissier, Olivier},
  _editor   = {Baroglio, Cristina and Hübner, Jomi Fred and Winikoff, Michael},
  year      = {2020},
  pages     = {42--60}
}

@inproceedings{RaoG95,
  author     = {Anand S. Rao and
                Michael P. Georgeff},
  _editor    = {Victor R. Lesser and
                Les Gasser},
  title      = {{BDI} Agents: From Theory to Practice},
  _booktitle = {Proceedings of the First International Conference on Multiagent Systems,
                June 12-14, 1995, San Francisco, California, {USA}},
  booktitle  = {Proceedings of the First International Conference on Multiagent Systems},
  pages      = {312--319},
  _publisher = {The {MIT} Press},
  year       = {1995},
  timestamp  = {Tue, 16 Nov 2004 11:26:50 +0100},
  biburl     = {https://dblp.org/rec/conf/icmas/RaoG95.bib},
  bibsource  = {dblp computer science bibliography, https://dblp.org}
}

@incollection{HubnerB09,
  author     = {Jomi Fred H{\"{u}}bner and
                Rafael H. Bordini},
  _editor    = {Adelinde M. Uhrmacher and
                Danny Weyns},
  title      = {Agent-Based Simulation Using {BDI} Programming in Jason},
  booktitle  = {Multi-Agent Systems - Simulation and Applications},
  _series    = {Computational Analysis, Synthesis, and Design of Dynamic Systems},
  pages      = {451--476},
  _publisher = {{CRC} Press / Taylor {\&} Francis},
  year       = {2009},
  _url       = {https://doi.org/10.1201/9781420070248.ch15},
  doi        = {10.1201/9781420070248.CH15},
  timestamp  = {Mon, 15 Jul 2019 08:16:18 +0200},
  biburl     = {https://dblp.org/rec/books/tf/09/HubnerB09.bib},
  bibsource  = {dblp computer science bibliography, https://dblp.org}
}

@inproceedings{nguyen_testing_2011,
  address    = {Berlin, Heidelberg},
  title      = {Testing in {Multi}-{Agent} {Systems}},
  isbn       = {978-3-642-19208-1},
  doi        = {10.1007/978-3-642-19208-1_13},
  language   = {en},
  booktitle  = {Agent-{Oriented} {Software} {Engineering} {X}},
  _publisher = {Springer},
  author     = {Nguyen, Cu D. and Perini, Anna and Bernon, Carole and Pavón, Juan and Thangarajah, John},
  _editor    = {Gleizes, Marie-Pierre and Gomez-Sanz, Jorge J.},
  year       = {2011},
  keywords   = {Autonomous Agent, Electronic Institution, Multiagent System, Testing Approach, Testing Level},
  pages      = {180--190}
}

@article{cossentino_passim_2008,
  title      = {{PASSIM}: a simulation-based process for the development of multi-agent systems},
  volume     = {2},
  issn       = {1746-1375},
  shorttitle = {{PASSIM}},
  _url       = {https://www.inderscienceonline.com/doi/abs/10.1504/IJAOSE.2008.017313},
  doi        = {10.1504/IJAOSE.2008.017313},
  number     = {2},
  urldate    = {2024-03-28},
  journal    = {International Journal of Agent-Oriented Software Engineering},
  author     = {Cossentino, Massimo and Fortino, Giancarlo and Garro, Alfredo and Mascillaro, Samuele and Wilma Russo},
  month      = jan,
  year       = {2008},
  note       = {Publisher: Inderscience Publishers},
  keywords   = {agent implementation, agent simulation, agent specification, agent-based systems, agent-oriented software engineering, AOSE, discrete-event simulation, e-marketplace, MASs, method engineering, multi-agent systems},
  pages      = {132--170}
}

@inproceedings{uhrmacher_simulation_2002,
  address = {San Antonio, TX, USA},
  title   = {Simulation for agent-oriented software engineering},
  url     = {https://pdfs.semanticscholar.org/b022/be67e1b2ff4f1f482ce3e4352608101ebcd8.pdf},
  urldate = {2024-03-28},
  author  = {Uhrmacher, Adelinde M.},
  year    = {2002}
}

@inproceedings{miles_why_2010,
  title     = {Why testing autonomous agents is hard and what can be done about it},
  url       = {https://www.academia.edu/download/69061967/miles.pdf},
  urldate   = {2024-03-28},
  booktitle = {{AOSE} {Technical} {Forum}},
  author    = {Miles, Simon and Winikoff, Michael and Cranefield, Stephen and Nguyen, Cu D. and Perini, Anna and Tonella, Paolo and Harman, Mark and Luck, Michael},
  year      = {2010}
}

@inproceedings{sakellariou_enhancing_2008,
  address   = {Berlin, Heidelberg},
  title     = {Enhancing {NetLogo} to Simulate {BDI} Communicating Agents},
  isbn      = {978-3-540-87881-0},
  doi       = {10.1007/978-3-540-87881-0_24},
  booktitle = {Artificial Intelligence: Theories, Models and Applications},
  publisher = {Springer},
  author    = {Sakellariou, Ilias and Kefalas, Petros and Stamatopoulou, Ioanna},
  editor    = {Darzentas, John and Vouros, George A. and Vosinakis, Spyros and Arnellos, Argyris},
  year      = {2008},
  pages     = {263–275},
  language  = {en}
}

@book{BordiniHW2007,
  title     = {Programming multi-agent systems in AgentSpeak using Jason},
  author    = {Bordini, Rafael H and H{\"u}bner, Jomi Fred and Wooldridge, Michael},
  year      = {2007},
  publisher = {John Wiley \& Sons},
  address   = {Oxford, England}
}

@inproceedings{BaiardiBCPOR24b,
  author    = {Martina Baiardi and
               Samuele Burattini and
               Giovanni Ciatto and
               Danilo Pianini and
               Alessandro Ricci and
               Andrea Omicini},
  editor    = {Daniela Briola and
               Rafael C. Cardoso and
               Brian Logan},
  title     = {On the External Concurrency of Current {BDI} Frameworks for {MAS}},
  booktitle = {Engineering Multi-Agent Systems - 12th International Workshop, {EMAS}
               2024, Revised Selected Papers},
  series    = {Lecture Notes in Computer Science},
  volume    = {15152},
  pages     = {42--63},
  publisher = {Springer},
  year      = {2024},
  address   = {Auckland, New Zealand},
  _url      = {https://doi.org/10.1007/978-3-031-71152-7\_3},
  doi       = {10.1007/978-3-031-71152-7\_3},
  timestamp = {Sat, 30 Nov 2024 21:10:38 +0100},
  biburl    = {https://dblp.org/rec/conf/emas/BaiardiBCPRO24.bib},
  bibsource = {dblp computer science bibliography, https://dblp.org}
}

@inproceedings{BaiardiBCPOR24,
  author     = {Martina Baiardi and
                Samuele Burattini and
                Giovanni Ciatto and
                Danilo Pianini and
                Andrea Omicini and
                Alessandro Ricci},
  _editor    = {Mehdi Dastani and
                Jaime Sim{\~{a}}o Sichman and
                Natasha Alechina and
                Virginia Dignum},
  title      = {Concurrency Model of {BDI} Programming Frameworks: Why Should We Control It?},
  _booktitle = {Proceedings of the 23rd International Conference on Autonomous Agents
                and Multiagent Systems, {AAMAS} 2024, Auckland, New Zealand, May 6-10,
                2024},
  booktitle  = {Proceedings {AAMAS} 2024},
  pages      = {2147--2149},
  _publisher = {{ACM}},
  year       = {2024},
  _url       = {https://dl.acm.org/doi/10.5555/3635637.3663089},
  doi        = {10.5555/3635637.3663089},
  timestamp  = {Fri, 03 May 2024 14:31:38 +0200},
  biburl     = {https://dblp.org/rec/conf/atal/BaiardiBCPOR24.bib},
  bibsource  = {dblp computer science bibliography, https://dblp.org}
}

@article{weyns2007aamas,
  author    = {Danny Weyns and
               Andrea Omicini and
               James Odell},
  title     = {Environment as a first class abstraction in multiagent systems},
  journal   = {Auton. Agents Multi Agent Syst.},
  volume    = {14},
  number    = {1},
  pages     = {5--30},
  year      = {2007},
  _url      = {https://doi.org/10.1007/s10458-006-0012-0},
  doi       = {10.1007/S10458-006-0012-0},
  timestamp = {Tue, 29 Dec 2020 18:21:00 +0100},
  biburl    = {https://dblp.org/rec/journals/aamas/WeynsOO07.bib},
  bibsource = {dblp computer science bibliography, https://dblp.org}
}

@inproceedings{pantoja2016atal,
  author    = {Carlos Eduardo Pantoja and
               Marcio Fernando Stabile Jr. and
               Nilson Mori Lazarin and
               Jaime Sim{\~{a}}o Sichman},
  editor    = {Matteo Baldoni and
               J{\"{o}}rg P. M{\"{u}}ller and
               Ingrid Nunes and
               Rym Zalila{-}Wenkstern},
  title     = {{ARGO:} An Extended Jason Architecture that Facilitates Embedded Robotic
               Agents Programming},
  booktitle = {{EMAS} 2016},
  address   = {Singapore},
  series    = {Lecture Notes in Computer Science},
  volume    = {10093},
  pages     = {136--155},
  publisher = {Springer},
  year      = {2016},
  _url      = {https://doi.org/10.1007/978-3-319-50983-9_8},
  doi       = {10.1007/978-3-319-50983-9_8},
  timestamp = {Sun, 02 Oct 2022 15:55:11 +0200},
  biburl    = {https://dblp.org/rec/conf/atal/PantojaSLS16.bib},
  bibsource = {dblp computer science bibliography, https://dblp.org}
}

@inproceedings{moro2022paams,
  author    = {Devis Dal Moro and
               Marco Robol and
               Marco Roveri and
               Paolo Giorgini},
  editor    = {Frank Dignum and
               Philippe Mathieu and
               Juan Manuel Corchado and
               Fernando de la Prieta},
  title     = {Developing BDI-Based Robotic Systems with {ROS2}},
  booktitle = {{PAAMS} 2022, Proceedings},
  series    = {Lecture Notes in Computer Science},
  volume    = {13616},
  pages     = {100--111},
  publisher = {Springer},
  address   = {L'Aquila, Italy},
  year      = {2022},
  _url      = {https://doi.org/10.1007/978-3-031-18192-4_9},
  doi       = {10.1007/978-3-031-18192-4_9},
  timestamp = {Mon, 05 Dec 2022 13:35:49 +0100},
  biburl    = {https://dblp.org/rec/conf/paams/MoroRRG22.bib},
  bibsource = {dblp computer science bibliography, https://dblp.org}
}

@inproceedings{brooks1991ijcai,
  author    = {Rodney A. Brooks},
  editor    = {John Mylopoulos and
               Raymond Reiter},
  title     = {Intelligence Without Reason},
  booktitle = {Proceedings of the 12th International Joint Conference on Artificial Intelligence},
  pages     = {569--595},
  publisher = {Morgan Kaufmann},
  address   = {Sydney, Australia},
  year      = {1991},
  month     = {August},
  url       = {http://ijcai.org/Proceedings/91-1/Papers/089.pdf},
  bibsource = {dblp computer science bibliography, https://dblp.org}
}

@article{o1998fipa,
  title     = {FIPA—towards a standard for software agents},
  author    = {O'Brien, Paul D and Nicol, Richard C},
  journal   = {BT Technology Journal},
  volume    = {16},
  pages     = {51--59},
  year      = {1998},
  publisher = {Springer}
}

@inproceedings{TaillandierBCAG16,
  author    = {Patrick Taillandier and
               Mathieu Bourgais and
               Philippe Caillou and
               Carole Adam and
               Benoit Gaudou},
  editor    = {Luis G. Nardin and
               Luis Antunes},
  title     = {A {BDI} Agent Architecture for the {GAMA} Modeling and Simulation
               Platform},
  booktitle = {Multi-Agent Based Simulation {XVII} - International Workshop, {MABS}
               2016, Revised Selected Papers},
  series    = {Lecture Notes in Computer Science},
  address   = {Singapore},
  month     = {May},
  volume    = {10399},
  pages     = {3--23},
  publisher = {Springer},
  year      = {2016},
  _url      = {https://doi.org/10.1007/978-3-319-67477-3_1},
  doi       = {10.1007/978-3-319-67477-3_1},
  timestamp = {Fri, 09 Apr 2021 18:47:26 +0200},
  biburl    = {https://dblp.org/rec/conf/mabs/TaillandierBCAG16.bib},
  bibsource = {dblp computer science bibliography, https://dblp.org}
}

@inproceedings{DrogoulACGGMTVVZ13,
  author    = {Alexis Drogoul and
               Edouard Amouroux and
               Philippe Caillou and
               Benoit Gaudou and
               Arnaud Grignard and
               Nicolas Marilleau and
               Patrick Taillandier and
               Maroussia Vavasseur and
               Duc{-}An Vo and
               Jean{-}Daniel Zucker},
  editor    = {Yves Demazeau and
               Toru Ishida and
               Juan M. Corchado and
               Javier Bajo},
  title     = {{GAMA:} {A} Spatially Explicit, Multi-level, Agent-Based Modeling
               and Simulation Platform},
  booktitle = {Advances on Practical Applications of Agents and Multi-Agent Systems,
               11th International Conference, {PAAMS} 2013},
  address   = {Salamanca, Spain},
  month     = {May},
  volume    = {7879},
  pages     = {271--274},
  publisher = {Springer},
  year      = {2013},
  _url      = {https://doi.org/10.1007/978-3-642-38073-0_25},
  doi       = {10.1007/978-3-642-38073-0_25},
  timestamp = {Thu, 20 Sep 2018 16:01:24 +0200},
  biburl    = {https://dblp.org/rec/conf/paams/DrogoulACGGMTVVZ13.bib},
  bibsource = {dblp computer science bibliography, https://dblp.org}
}

@inproceedings{uhrmacher1998agents,
  title        = {Agents in discrete event simulation},
  author       = {Uhrmacher, Adelinde M. and Schattenberg, Bernd},
  booktitle    = {European Simulation Symposium-ESS},
  volume       = {98},
  pages        = {129--136},
  year         = {1998},
  organization = {Citeseer}
}

@article{caballero2011eaai,
  author    = {Alberto Caballero and
               Juan A. Bot{\'{\i}}a and
               Antonio Fernandez G{\'{o}}mez{-}Skarmeta},
  title     = {Using cognitive agents in social simulations},
  journal   = {Eng. Appl. Artif. Intell.},
  volume    = {24},
  number    = {7},
  pages     = {1098--1109},
  year      = {2011},
  _url      = {https://doi.org/10.1016/j.engappai.2011.06.006},
  doi       = {10.1016/J.ENGAPPAI.2011.06.006},
  timestamp = {Tue, 21 Mar 2023 21:06:24 +0100},
  biburl    = {https://dblp.org/rec/journals/eaai/CaballeroBG11.bib},
  bibsource = {dblp computer science bibliography, https://dblp.org}
}

@article{SNCS:JaktaJournal,
  author  = {Martina Baiardi and
             Samuele Burattini and
             Giovanni Ciatto and
             Danilo Pianini},
  title   = {Blending {BDI} agents with object-oriented and functional programming with {JaKtA}},
  journal = {{SN} Computer Science},
  year    = {2025},
  _url    = {https://doi.org/10.1007/s42979-024-03244-y},
  doi     = {10.1007/s42979-024-03244-y}
}

@book{simchallenges,
  author    = {Fujimoto, Richard and Bock, Conrad and Chen, Wei and Page, Ernest and Panchal, Jitesh H.},
  title     = {Research Challenges in Modeling and Simulation for Engineering Complex Systems},
  isbn      = {9783319585444},
  issn      = {2195-2825},
  _url      = {http://dx.doi.org/10.1007/978-3-319-58544-4},
  doi       = {10.1007/978-3-319-58544-4},
  journal   = {Simulation Foundations,  Methods and Applications},
  publisher = {Springer International Publishing},
  address   = {Berlin},
  year      = {2017}
}

@article{Cai2007,
  title     = {Exact stochastic simulation of coupled chemical reactions with delays},
  volume    = {126},
  issn      = {1089-7690},
  _url      = {http://dx.doi.org/10.1063/1.2710253},
  doi       = {10.1063/1.2710253},
  number    = {12},
  journal   = {The Journal of Chemical Physics},
  publisher = {AIP Publishing},
  author    = {Cai,  Xiaodong},
  year      = {2007},
  month     = mar
}

@article{Richardson2000,
  title     = {Direct Large-Scale N-Body Simulations of Planetesimal Dynamics},
  volume    = {143},
  issn      = {0019-1035},
  _url      = {http://dx.doi.org/10.1006/icar.1999.6243},
  doi       = {10.1006/icar.1999.6243},
  number    = {1},
  journal   = {Icarus},
  publisher = {Elsevier BV},
  author    = {Richardson,  D},
  year      = {2000},
  month     = jan,
  pages     = {45–59}
}

@inproceedings{Edmonds00,
  author    = {Bruce Edmonds},
  editor    = {Scott Moss and
               Paul Davidsson},
  title     = {The Use of Models - Making {MABS} More Informative},
  booktitle = {Multi-Agent-Based Simulation, Second International Workshop, {MABS}
               2000, Revised and Additional Papers},
  series    = {Lecture Notes in Computer Science},
  volume    = {1979},
  pages     = {15--32},
  publisher = {Springer},
  address   = {Boston, MA, USA},
  month     = {July},
  year      = {2000},
  _url      = {https://doi.org/10.1007/3-540-44561-7_2},
  doi       = {10.1007/3-540-44561-7_2},
  timestamp = {Wed, 25 Sep 2019 18:15:39 +0200},
  biburl    = {https://dblp.org/rec/conf/mabs/Edmonds00.bib},
  bibsource = {dblp computer science bibliography, https://dblp.org}
}

@incollection{DBLP:books/tf/09/DrogoulMF09,
  author    = {Alexis Drogoul and
               Fabien Michel and
               Jacques Ferber},
  editor    = {Adelinde M. Uhrmacher and
               Danny Weyns},
  title     = {Multi-Agent Systems and Simulation: {A} Survey from the Agent Community's
               Perspective},
  booktitle = {Multi-Agent Systems - Simulation and Applications},
  series    = {Computational Analysis, Synthesis, and Design of Dynamic Systems},
  pages     = {3--51},
  publisher = {{CRC} Press / Taylor {\&} Francis},
  year      = {2009},
  address   = {Boca Raton, FL, USA},
  _url      = {https://doi.org/10.1201/9781420070248.pt1},
  doi       = {10.1201/9781420070248.PT1},
  timestamp = {Fri, 12 Jul 2019 11:06:16 +0200},
  biburl    = {https://dblp.org/rec/books/tf/09/DrogoulMF09.bib},
  bibsource = {dblp computer science bibliography, https://dblp.org}
}

@inproceedings{DBLP:conf/ecal/JacobiHH95,
  author    = {Nick Jakobi and
               Phil Husbands and
               Inman Harvey},
  editor    = {Federico Mor{\'{a}}n and
               Alvaro Moreno and
               Juan Juli{\'{a}}n Merelo Guerv{\'{o}}s and
               Pablo Chac{\'{o}}n},
  title     = {Noise and the Reality Gap: The Use of Simulation in Evolutionary Robotics},
  booktitle = {Advances in Artificial Life, Third European Conference on Artificial Life},
  series    = {Lecture Notes in Computer Science},
  volume    = {929},
  pages     = {704--720},
  publisher = {Springer},
  address   = {Granada, Spain},
  month     = {June},
  year      = {1995},
  _url      = {https://doi.org/10.1007/3-540-59496-5_337},
  doi       = {10.1007/3-540-59496-5_337},
  timestamp = {Sun, 06 Oct 2024 21:00:03 +0200},
  biburl    = {https://dblp.org/rec/conf/ecal/JacobiHH95.bib},
  bibsource = {dblp computer science bibliography, https://dblp.org}
}

@article{DBLP:journals/tec/KoosMD13,
  author    = {Sylvain Koos and
               Jean{-}Baptiste Mouret and
               St{\'{e}}phane Doncieux},
  title     = {The Transferability Approach: Crossing the Reality Gap in Evolutionary
               Robotics},
  journal   = {{IEEE} Trans. Evol. Comput.},
  volume    = {17},
  number    = {1},
  pages     = {122--145},
  year      = {2013},
  _url      = {https://doi.org/10.1109/TEVC.2012.2185849},
  doi       = {10.1109/TEVC.2012.2185849},
  timestamp = {Tue, 12 May 2020 16:50:56 +0200},
  biburl    = {https://dblp.org/rec/journals/tec/KoosMD13.bib},
  bibsource = {dblp computer science bibliography, https://dblp.org}
}

@inproceedings{DBLP:conf/wsc/MoonH13,
  author    = {Il{-}Chul Moon and
               Jeong{-}Hee Hong},
  title     = {Theoretic interplay between abstraction, resolution, and fidelity
               in model information},
  booktitle = {Winter Simulations Conference: Simulation Making Decisions in a Complex
               World, {WSC} 2013},
  pages     = {1283--1291},
  publisher = {{IEEE}},
  address   = {Washington, DC, USA},
  month     = {December},
  year      = {2013},
  _url      = {https://doi.org/10.1109/WSC.2013.6721515},
  doi       = {10.1109/WSC.2013.6721515},
  timestamp = {Thu, 10 Jun 2021 22:20:28 +0200},
  biburl    = {https://dblp.org/rec/conf/wsc/MoonH13.bib},
  bibsource = {dblp computer science bibliography, https://dblp.org}
}

@article{DBLP:journals/csse/FortinoGR05,
  author    = {Giancarlo Fortino and
               Alfredo Garro and
               Wilma Russo},
  title     = {An integrated approach for the development and validation of multi-agent
               systems},
  journal   = {Comput. Syst. Sci. Eng.},
  volume    = {20},
  number    = {4},
  year      = {2005},
  bibsource = {dblp computer science bibliography, https://dblp.org}
}

@article{DBLP:journals/jos/FortinoN13,
  author    = {Giancarlo Fortino and
               Michael J. North},
  title     = {Simulation-based development and validation of multi-agent systems:
               {AOSE} and {ABMS} approaches},
  journal   = {J. Simulation},
  volume    = {7},
  number    = {3},
  pages     = {137--143},
  year      = {2013},
  _url      = {https://doi.org/10.1057/jos.2013.12},
  doi       = {10.1057/JOS.2013.12},
  timestamp = {Tue, 06 Jun 2017 22:22:01 +0200},
  biburl    = {https://dblp.org/rec/journals/jos/FortinoN13.bib},
  bibsource = {dblp computer science bibliography, https://dblp.org}
}

@article{computers10020016,
  author         = {Cardoso, Rafael C. and Ferrando, Angelo},
  title          = {A Review of Agent-Based Programming for Multi-Agent Systems},
  journal        = {Computers},
  volume         = {10},
  year           = {2021},
  number         = {2},
  article-number = {16},
  issn           = {2073-431X},
  doi            = {10.3390/computers10020016}
}

@article{Taillandier2018,
  title     = {Building,  composing and experimenting complex spatial models with the GAMA platform},
  volume    = {23},
  issn      = {1573-7624},
  doi       = {10.1007/s10707-018-00339-6},
  number    = {2},
  journal   = {GeoInformatica},
  publisher = {Springer Science and Business Media LLC},
  author    = {Taillandier,  Patrick and Gaudou,  Benoit and Grignard,  Arnaud and Huynh,  Quang-Nghi and Marilleau,  Nicolas and Caillou,  Philippe and Philippon,  Damien and Drogoul,  Alexis},
  year      = {2018},
  month     = dec,
  pages     = {299–322}
}

@inproceedings{DBLP:conf/coordination/AguzziBBCCDFPV25,
    author       = {Gianluca Aguzzi and
                  Lorenzo Bacchini and
                  Martina Baiardi and
                  Roberto Casadei and
                  Angela Cortecchia and
                  Davide Domini and
                  Nicolas Farabegoli and
                  Danilo Pianini and
                  Mirko Viroli},
    editor       = {Cinzia Di Giusto and
                  Ant{\'{o}}nio Ravara},
    title        = {A Demonstrator for Self-organizing Robot Teams},
    booktitle    = {Coordination Models and Languages - 27th {IFIP} {WG} 6.1 International
                  Conference, {COORDINATION} 2025, Held as Part of the 20th International
                  Federated Conference on Distributed Computing Techniques, DisCoTec
                  2025, Lille, France, June 17-19, 2025, Proceedings},
    series       = {Lecture Notes in Computer Science},
    address = {Lille, France},
    volume       = {15731},
    pages        = {230--244},
    publisher    = {Springer},
    year         = {2025},
    _url          = {https://doi.org/10.1007/978-3-031-95589-1\_12},
    doi          = {10.1007/978-3-031-95589-1\_12},
    timestamp    = {Fri, 04 Jul 2025 22:05:42 +0200},
    biburl       = {https://dblp.org/rec/conf/coordination/AguzziBBCCDFPV25.bib},
    bibsource    = {dblp computer science bibliography, https://dblp.org}
}

@inproceedings{DBLP:conf/dsrt/Baiardi24,
  author       = {Martina Baiardi},
  title        = {Validation of {BDI} MASs via Simulation},
  booktitle    = {28th International Symposium on Distributed Simulation and Real Time
                  Applications, {DS-RT} 2024, Urbino, Italy, October 7-9, 2024},
  pages        = {128--129},
  publisher    = {{IEEE}},
  year         = {2024},
  _url          = {https://doi.org/10.1109/DS-RT62209.2024.00029},
  doi          = {10.1109/DS-RT62209.2024.00029},
  timestamp    = {Thu, 17 Apr 2025 16:07:35 +0200},
  biburl       = {https://dblp.org/rec/conf/dsrt/Baiardi24.bib},
  bibsource    = {dblp computer science bibliography, https://dblp.org},
  address = {Urbino, Italy}
}

@inbook{Briola2025,
  title = {Enhancing Testing of MASs with a Simulator of JADE},
  ISBN = {9781643686318},
  ISSN = {1879-8314},
  _url = {http://dx.doi.org/10.3233/FAIA251247},
  DOI = {10.3233/faia251247},
  booktitle = {ECAI 2025},
  publisher = {IOS Press},
  author = {Briola,  Daniela and Vizzari,  Giuseppe and Montinaro,  Mauro},
  year = {2025},
  pages = {3687 - 3694},
  month = oct,
  editor    = {Lynce, In{\^e}s and Murano, Nello and Vallati, Mauro 
               and Villata, Serena and Chesani, Federico and Milano, Michela 
               and Omicini, Andrea and Dastani, Mehdi},
  address = {Amsterdam, Netherlands}
}

@book{DBLP:books/wi/BellifemineCG07,
  author       = {Fabio Bellifemine and
                  Giovanni Caire and
                  Dominic Greenwood},
  title        = {Developing Multi-Agent Systems with {JADE}},
  address = {Chichester, UK},
  publisher    = {Wiley},
  year         = {2007},
  _url          = {https://doi.org/10.1002/9780470058411},
  doi          = {10.1002/9780470058411},
  isbn         = {978-0-47005747-6},
  timestamp    = {Tue, 23 Jul 2019 16:10:34 +0200},
  biburl       = {https://dblp.org/rec/books/wi/BellifemineCG07.bib},
  bibsource    = {dblp computer science bibliography, https://dblp.org}
}

@article{sociotech,
    author       = {Gordon D. Baxter and
                  Ian Sommerville},
    title        = {Socio-technical systems: From design methods to systems engineering},
    journal      = {Interact. Comput.},
    volume       = {23},
    number       = {1},
    pages        = {4--17},
    year         = {2011},
    _url          = {https://doi.org/10.1016/j.intcom.2010.07.003},
    doi          = {10.1016/J.INTCOM.2010.07.003},
    timestamp    = {Fri, 13 Mar 2020 10:53:27 +0100},
    biburl       = {https://dblp.org/rec/journals/iwc/BaxterS11.bib},
    bibsource    = {dblp computer science bibliography, https://dblp.org}
}

@ARTICLE{cps,
    author={Guan, Xinping and Yang, Bo and Chen, Cailian and Dai, Wenbin and Wang, Yiyin},
    journal={IEEE/CAA Journal of Automatica Sinica},
    title={A comprehensive overview of cyber-physical systems: from perspective of feedback system},
    year={2016},
    volume={3},
    number={1},
    pages={1-14},
    doi={10.1109/JAS.2016.7373757}}

@article{pervasive,
    author       = {Panos E. Kourouthanassis and
                  George M. Giaglis and
                  Dimitrios C. Karaiskos},
    title        = {Delineating 'pervasiveness' in pervasive information systems: a taxonomical
                  framework and design implications},
    journal      = {J. Inf. Technol.},
    volume       = {25},
    number       = {3},
    pages        = {273--287},
    year         = {2010},
    _url          = {https://doi.org/10.1057/jit.2009.6},
    doi          = {10.1057/JIT.2009.6},
    timestamp    = {Tue, 07 May 2024 20:28:12 +0200},
    biburl       = {https://dblp.org/rec/journals/jitech/KourouthanassisGK10.bib},
    bibsource    = {dblp computer science bibliography, https://dblp.org}
}

@article{DBLP:journals/tosem/ParryKHM22,
    author       = {Owain Parry and
                  Gregory M. Kapfhammer and
                  Michael Hilton and
                  Phil McMinn},
    title        = {A Survey of Flaky Tests},
    journal      = {{ACM} Trans. Softw. Eng. Methodol.},
    volume       = {31},
    number       = {1},
    pages        = {17:1--17:74},
    year         = {2022},
    _url          = {https://doi.org/10.1145/3476105},
    doi          = {10.1145/3476105},
    timestamp    = {Sun, 12 Nov 2023 02:19:58 +0100},
    biburl       = {https://dblp.org/rec/journals/tosem/ParryKHM22.bib},
    bibsource    = {dblp computer science bibliography, https://dblp.org}
}

\end{document}